\documentclass{emulateapj}

\shorttitle{SSCs in SBS 0335-052}
\shortauthors{Reines, Johnson, \& Hunt}

\begin{document}
 
\title{A New View of the Super Star Clusters in the Low-Metallicity Galaxy SBS 0335-052}

\author{Amy E. Reines and Kelsey E. Johnson\footnote{Adjunct at the National Radio
Astronomy Observatory, 520 Edgemont Road, Charlottesville, VA 22903, USA}}
\affil{Department of Astronomy, University of Virginia,
Charlottesville, VA, 22904-4325}
\email{areines@virginia.edu, kej7a@virginia.edu}

\and

\author{Leslie K. Hunt}
\affil{INAF-Istituto di Radioastronomia-Sez. Firenze, L.go, Fermi 5,
I-50125 Firenze, Italy}
\email{hunt@arcetri.astro.it}

\begin{abstract}

We present a study of the individual super star clusters (SSCs) in the 
low-metallicity galaxy SBS 0335-052 using new near-infrared and archival
optical {\it Hubble Space Telescope} observations.  The physical properties
of the SSCs are derived from fitting model spectral energy distributions (SEDs)
to the optical photometry, as well as from the H$\alpha$ and Pa$\alpha$
nebular emission.  Among the clusters, we find a significant age spread that
is correlated with position in the galaxy, suggesting successive cluster formation
occurred in SBS 0335-052 triggered by a large-scale disturbance traveling through
the galaxy at a speed of $\sim 35$ km s$^{-1}$.  The SSCs exhibit $I$-band
($\sim 0.8~\mu$m) and near-IR ($\sim 1.6-2.1~\mu$m) excesses with respect to model
SEDs fit to the optical data.  We hypothesize that the $I$-band excess is dominated
by a photoluminescent process known as Extended Red Emission; however, this mechanism
cannot account for the excesses observed at longer near-IR wavelengths.  From the cluster SEDs
and colors, we find that the primary origin of the near-IR excess observed in the youngest
SSCs ($\lesssim 3$ Myr) is hot dust emission, while evolved red supergiants
dominate the near-IR light in the older ($\gtrsim 7$ Myr) clusters.  We also
find evidence for a porous and clumpy interstellar medium (ISM) surrounding the youngest,
embedded SSCs: the ionized gas emission underpredicts the expected ionizing
luminosities from the optical stellar continuum, suggesting ionizing photons
are leaking out of the immediate vicinity of the clusters before ionizing hydrogen.
The corrected, intrinsic ionizing luminosities of the two SSCs younger than $\sim 3$~Myr
are each $\sim 5 \times 10^{52}~{\rm s}^{-1}$, which is equivalent to each cluster hosting
$\sim 5000$ O7.5 V stars.  The inferred masses of these SSCs are $\sim 10^6 M_\odot$.

\end{abstract}


\section{Introduction}

Ancient globular clusters are among the oldest objects known, almost
as old as the universe itself with ages estimated in excess of
12~Gyr \citep{Vandenberg96, Freeman02}.  As such, these clusters are
valuable relics of the earlier universe, when violent and intense
star formation was common.  Indeed, globular clusters are ubiquitous
around massive galaxies today \citep{Harris91, Brodie06}, and they must have
been formed prodigiously in the primordial universe given that
$\gtrsim 90\%$ may have been subsequently destroyed \citep{Fall01,Whitmore07}.

However, the conditions required for the creation of globular clusters have puzzled
astronomers for decades. For many years the prevailing belief was that
globular clusters were simply formed by the gravitational collapse of
density inhomogeneities in the early universe \citep{Peebles68, Fall85},
and little was known about their early evolution. However, in the mid
1990's, observations using the {\it Hubble Space Telescope}
discovered extremely young, massive, and compact clusters in the local
universe, the so-called ``super star clusters'' (SSCs) that have
properties consistent with those expected for adolescent globular
clusters \citep[e.g.][]{Whitmore03}.  The discovery of SSCs precipitated a
major shift in our understanding of the conditions required for globular cluster
formation.

Although the evidence connecting young SSCs and ancient globular clusters
is compelling, a critical issue remains: modern day SSCs and
ancient globular clusters were formed in environments with very
different metallicities.  We have not yet observationally constrained
what effect low metal abundances had on the formation of massive star clusters.
Much effort has gone into understanding the formation of the {\it first} stars in
the universe \citep[e.g. those stars formed out of truly primordial
material,][]{Bromm03,Santoro06,Tumlinson07}, and it is clear that
metallicity has a critical role in this regime.  Even moderately low
metal abundances may affect star formation in a number of ways,
including dust formation, cooling and pressure, and the hardness of
the resulting stellar spectra.

\begin{figure*}[!t]
\begin{center}
\includegraphics[scale=0.61]{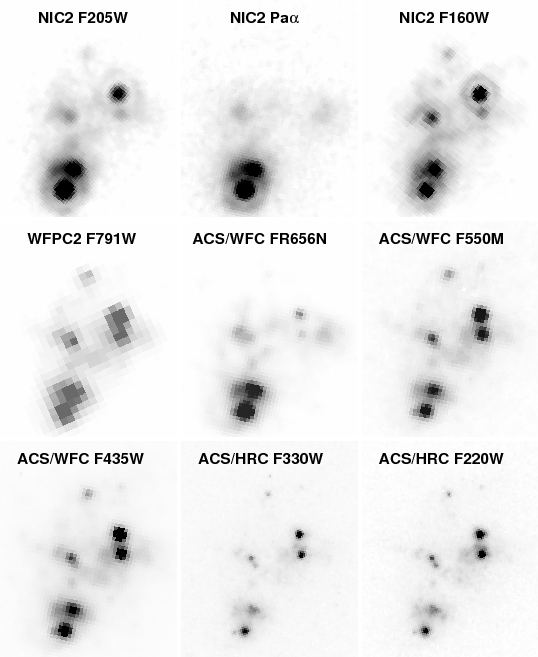}
\caption{{\it HST} images of the observations used in this work.
\label{allims}}
\end{center}
\end{figure*}

In an effort to investigate massive star cluster formation in an
environment similar to that which might be found in primordial
galaxies during the time ancient globular clusters were prolifically
formed, we have targeted the galaxy SBS~0335-052 (E) for a detailed
multi-wavelength study.  SBS~0335-052 is a remarkable blue compact
dwarf galaxy at a distance of 54 Mpc (NED\footnote{NASA/IPAC Extragalactic Database})
that is well-known for its extremely low oxygen abundance of $\sim 1/40Z_\odot$
\citep{Izotov90,Melnick92,Izotov01}.  Unlike I~Zw~18 and SBS 0335-052W, which have
slightly lower metallicities, SBS~0335-052 is undergoing a vigorous
starburst with a star formation rate $\gtrsim 1~M_\odot$~yr$^{-1}$ and
hosts extremely massive young clusters, making it ideal for this study
\citep{Thuan97, Thuan99, Hunt01, Dale01, Plante02, Hunt04, Houck04}.

Here we present new near-infrared and archival optical high-resolution
observations of SBS 0335-052 with the goal of studying SSC
formation and evolution at low metallicity.  Near-IR observations are
particularly important for probing the dusty birth environments of
embedded massive star clusters, and the wavelength range of $\sim
1-2~\mu$m is well-suited to assessing a cluster's evolutionary state
because it samples the spectral energy distribution (SED) at the nexus of
emission between dust and stellar light.  

The {\it HST} observations and photometry are presented in \S\ref{hstsec} and
\S\ref{photsec}.  In \S\ref{propsec}, we discuss various properties of the SSCs.
Specifically, we derive the physical properties of the SSCs from fitting model SEDs
to the optical photometry, as well as from the H$\alpha$, Pa$\alpha$, and thermal
radio nebular emission \citep{Johnson08}.  In addition, we present evidence
for a porous and clumpy ISM in the youngest embedded clusters in SBS 0335-052
which can account for the apparently discrepant extinction estimates 
found in the literature.  We also provide explanations for the observed
{\it I}-band ($\sim 0.8~\mu$m) and near-IR ($\sim 1.6-2.1~\mu$m)
excesses.  Evidence for successive cluster formation is given in
\S\ref{suc_sec}, and finally, our conclusions are summarized in
\S\ref{conclusions}.

\section{{\it HST} Observations}\label{hstsec}

\begin{figure}[!t]
\begin{center}
\includegraphics[scale=0.44]{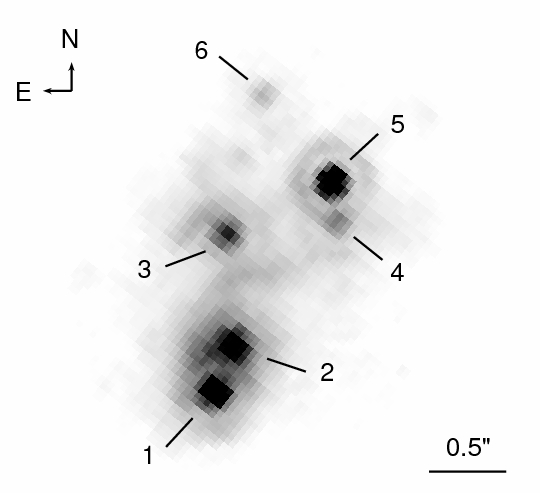}
\caption{NICMOS F160W ($\sim H$-band) image with the SSCs identified by
\citet{Thuan97}.
\label{IDim}}
\end{center}
\end{figure}

Multi-wavelength imaging of SBS~0335-052 has been obtained with
the {\it Hubble Space Telescope}\footnote{Based on
observations made with the NASA/ESA Hubble Space Telescope, obtained
at the Space Telescope Science Institute, which is operated by the
Association of Universities for Research in Astronomy, Inc., under
NASA contract NAS 5-26555. These observations are associated with
program \#10894, 10575, \& 5408.}.  New high-resolution near-IR
observations are presented along with archival UV and optical images
of the galaxy.  A summary of the observations used in this work is
given in Table~\ref{hstobs} and the images are shown in
Figure \ref{allims}.  The SSCs (1-6) are identified in Figure \ref{IDim}
using the notation of \citet{Thuan97}.

\begin{deluxetable}{cccc} 
\tabletypesize{\footnotesize}
\tablecolumns{4} 
\tablewidth{0pt} 
\tablecaption{{\it HST} Observations of SBS~0335-052\label{hstobs}} 
\tablehead{ 
\colhead{Filter}  &  \colhead{Instrument} & \colhead{Description} & \colhead{Exp. Time (s)}}
\startdata
\cutinhead{New Near-IR Observations\tablenotemark{a}}
F160W & NIC 2   & $1.6~\mu$m, $\sim$H   & 2431  \\ 
F187N & NIC 2   & Pa$\alpha$ continuum    & 4862  \\ 
F190N & NIC 2   & P$\alpha$ (redshifted) & 4606  \\ 
F205W & NIC 2   & $2.1~\mu$m, $\sim$K     & 2175  \\ 
\cutinhead{Archival UV and Optical Observations\tablenotemark{b}}
F220W & ACS/HRC & Near-UV       & 1660  \\
F330W & ACS/HRC & HRC U         & 800   \\
F435W & ACS/WFC & Johnson B     & 680   \\
F550M & ACS/WFC & Narrow V      & 430   \\
FR656N & ACS/WFC & H$\alpha$    & 680   \\
F791W & WFPC2   & I             & 4400 \\
\enddata
\tablenotetext{a}{The NIC 2 observations are associated with Proposal ID 10894 (this work).}
\tablenotetext{b}{The ACS and WFPC2 observations are associated with Proposal IDs 10575
(PI G. Ostlin) and 5408 (PI T. Thuan), respectively.}
\end{deluxetable} 

\subsection{Near-IR Imaging with NICMOS}

High-resolution near-IR observations of SBS 0335-052 were obtained
with the Near Infrared Camera and Multiobject Spectrometer (NICMOS)
on 2006 September 28 and 2006 December 25-26.  Using the NIC 2 Camera,
broad-band images at $1.6~\mu$m ($\sim H$) and $2.1~\mu$m ($\sim K$) were obtained
through the F160W and F205W filters, respectively, and narrow-band
Paschen $\alpha$ line and continuum images were obtained through
the F190N and F187N filters, respectively\footnote{At the redshift of
SBS 0335-052, the Pa$\alpha$ line falls in the F190N filter and the F187N
filter is used to obtain the continuum.}.  A spiral dither pattern
was implemented to allow improvement of the spatial resolution of the
observations by better sampling the point spread function (PSF) of the
NIC 2 Camera.  In addition, off-source observations were obtained
in the F187N, F190N, and F205W filters in order to mitigate the
effects of the thermal background.

\begin{figure*}
\begin{center}
\includegraphics[scale=0.8]{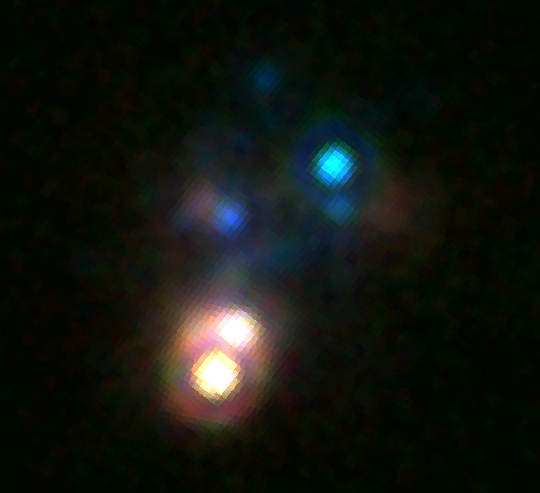}
\caption{NICMOS three-color image: F160W ($\sim H$) in blue, F205W ($\sim K$) in
green, and Pa$\alpha$ (continuum subtracted) in red.
\label{nicim}}
\end{center}
\end{figure*}

The raw data were processed with the {\tt calnica} NICMOS calibration
pipeline.  The calibrated NICMOS data were further reduced using a
combination of custom IDL programs and the Drizzle software
\citep{Fruchter02} available in the STSDAS\footnote{The Space Telescope
Science Data Analysis System} DITHER package in IRAF\footnote{The Image
Reduction and Analysis Facility is distributed by the National Optical
Astronomical Observatories, which are operated by the Association of Universities
for Research in Astronomy (AURA) Inc., under cooperative agreement with
the National Science Foundation (NSF).}.
The residual flat imprint and ``pedestal effect'' that can plague
NICMOS observations were effectively removed by subtracting off-source
``sky'' images from on-source images for each dither position\footnote{
This step was not applied to the F160W images since thermal
fluctuations are minimal at $1.6 \mu$m and we did not chop to sky for
these observations.} and then subtracting the quadrant-dependent bias
levels.  The resulting images (8 for each of the F160W and F205W filters
and 16 for each of the F187N and F190N filters) were then combined using
the Drizzle software to remove cosmic rays and improve the spatial
resolution of the images.  The drizzle parameter {\tt scale} was set to
0.5 to reduce the native NIC 2 plate scale of $\sim$0\farcs075 pixel$^{-1}$ to
$\sim$0\farcs0375 pixel$^{-1}$.  The final drizzled images are essentially
diffraction limited, with spatial resolutions in the range
$\sim$0\farcs15 ($1.6~\mu$m) to $\sim$0\farcs2 ($2.1~\mu$m).  
Figure \ref{nicim} shows a three-color NICMOS image:
F160W (blue), F205W (green), and Pa$\alpha$ (red). 

\subsection{Archival UV and Optical Images}

Optical and UV observations of SBS 0335-052 have been obtained with
the Advanced Camera for Surveys (ACS) and we have retrieved the
pipeline-produced calibrated and drizzled images from the archive.  Images through
the F220W ($\sim$~Near-UV) and F330W ($\sim U$) filters were obtained with
the High Resolution Channel (HRC) and images through the F435W ($\sim B$),
F550M ($\sim V$), and FR656N (H$\alpha$) filters were obtained with the Wide
Field Channel (WFC).  The HRC and WFC have plate scales of $\sim$0\farcs027
pixel$^{-1}$ and $\sim$0\farcs05 pixel$^{-1}$, respectively.  The HRC UV
images have a resolution of $\sim$0\farcs06 and the WFC optical images have
a resolution of $\sim$0\farcs125.   SSC 3 is actually
two sources at the resolution of the ACS/HRC.

Two images of SBS 0335-052 through the F791W ($\sim I$) filter were
obtained with the Wide Field and Planetary Camera 2 (WFPC2) and we combined
the pipeline-produced calibrated images from the archive and rejected cosmic
rays.  The galaxy was placed on the WF3 chip which has a plate scale of
$\sim$0\farcs1 pixel$^{-1}$.  The resolution of the F791W image is
$\sim$0\farcs2.

\section{Photometry of the Super Star Clusters in SBS 0335-052}\label{photsec}

We performed aperture photometry on the super star clusters
identified by \citet{Thuan97}
using a custom IDL photometry program (SURPHOT) allowing for
consistent apertures and background annuli across multiple wavebands.
A full description of SURPHOT is given in \citet{Reines08}.

The SSCs are identified in the F160W image \citep[using the notation of]
[see Figure \ref{IDim}]{Thuan97} and the flux density at each
wavelength is measured within a circular aperture of radius 0\farcs15.
The background contribution at each wavelength is determined
within an annulus of inner and outer radii equal to 0\farcs45 and 0\farcs65,
respectively.  Aperture corrections
are determined from simulated
PSFs for each camera-filter combination and position on the detector using the
Tiny Tim software package (v6.3) distributed by the Space Telescope Science
Institute \citep{Krist04}.  Final aperture-corrected flux densities are obtained
by multiplying the measured background-subtracted flux densities within a 0\farcs15
radius aperture by the following factors: 1.29(F220W), 1.21(F330W), 1.23(F435W),
1.22(F550M), 1.22(FR656N), 1.68(F791W), 1.85(F160W), 1.87(F187N), 1.88(F190N), and
1.94(F205W).

Final aperture-corrected flux densities and {\it HST} system magnitudes with Vega
zeropoints of the SSCs are listed in Table \ref{SSCphot}.  Pa$\alpha$ fluxes,
H$\alpha$ fluxes, and H$\alpha$ equivalent widths of the SSCs
are given in Table \ref{linetable}.  The width of the FR656N ($\sim$131 \AA)
filter is such that it would contain any [NII] emission in addition to H$\alpha$.  The
[NII] emission in SBS 0335-052, however, is negligible \citep{Izotov90}.
The H$\alpha$ equivalent widths are obtained by dividing the total (continuum
subtracted) H$\alpha$ flux in the FR656N filter by the continuum flux density
at the redshifted wavelength of H$\alpha$ (6651 \AA), which is found by
interpolating between the F550M and F791W filters.  We note that the Pa$\alpha$ flux
measured in the entire line-emitting region containing SSCs 1 and 2 is equal
to that found by \citet{Thompson06}, $1.77 \times 10^{-14}$ erg s$^{-1}$ cm$^{-2}$.
The H$\alpha$ flux in this region is $1.66 \times 10^{-13}$ erg s$^{-1}$ cm$^{-2}$.

\begin{deluxetable*}{ccccccccc}
\tabletypesize{\footnotesize}
\tablecolumns{9}
\tablewidth{0pt}
\tablecaption{Photometry of the SSCs in SBS 0335-052 \label{SSCphot}}
\tablehead{
\colhead{Source} & \colhead{F220W} & \colhead{F330W} & \colhead{F435W} &
\colhead{F550M} & \colhead{F791W} & \colhead{F160W} & \colhead{F187N} & \colhead{F205W}}
\startdata
\cutinhead{Log Flux Density (erg s$^{-1}$ cm$^{-2}$ \AA$^{-1}$ )}
SSC 1 &  -15.66 &  -15.90 &  -16.10 &  -16.51 &  -16.65 &  -17.56 &  -17.51 & 
-17.35 \\
SSC 2 &  -15.69 &  -15.98 &  -16.20 &  -16.58 &  -16.83 &  -17.60 &  -17.68 & 
-17.49 \\
SSC 3 &  -15.86 &  -16.28 &  -16.53 &  -16.86 &  -17.11 &  -17.88 &  -18.09 & 
-18.08 \\
SSC 4 &  -15.39 &  -15.87 &  -16.15 &  -16.55 &  -16.87 &  -18.06 &  -18.13 & 
-18.18 \\
SSC 5 &  -15.34 &  -15.78 &  -15.98 &  -16.27 &  -16.56 &  -17.58 &  -17.64 & 
-17.73 \\
SSC 6 &  -16.31 &  -16.71 &  -16.91 &  -17.18 &  -17.41 &  -18.37 &  -18.54 & 
-18.46 \\
\cutinhead{{\it HST} System Magnitude with Vega Zeropoint}
SSC 1 &  18.29 &  18.53 &  19.78 &  20.11 &  19.29 &  19.16 &  18.37 &  17.65 \\
SSC 2 &  18.37 &  18.73 &  20.02 &  20.28 &  19.75 &  19.27 &  18.79 &  18.01 \\
SSC 3 &  18.79 &  19.48 &  20.85 &  20.98 &  20.44 &  19.96 &  19.82 &  19.49 \\
SSC 4 &  17.62 &  18.45 &  19.90 &  20.20 &  19.85 &  20.41 &  19.92 &  19.75 \\
SSC 5 &  17.50 &  18.24 &  19.47 &  19.52 &  19.07 &  19.20 &  18.68 &  18.61 \\
SSC 6 &  19.91 &  20.54 &  21.80 &  21.79 &  21.20 &  21.17 &  20.95 &  20.43 \\
\enddata
\tablecomments{We adopt a distance modulus of 33.66 for SBS 0335-052.  The
uncertainties in the flux densities are $\sim 10\%$.}
\end{deluxetable*}

\begin{deluxetable*}{cccc}
\tabletypesize{\footnotesize}
\tablecolumns{4}
\tablewidth{0pt}
\tablecaption{Nebular Emission from the SSCs \label{linetable}}
\tablehead{
\colhead{Source} & \colhead{Pa$\alpha$ Flux ($\times 10^{-15}$} & 
\colhead{H$\alpha$ Flux ($\times 10^{-15}$} & \colhead{H$\alpha$ Equivalent} \\
\colhead{ } & \colhead{erg s$^{-1}$ cm$^{-2}$)} & 
\colhead{erg s$^{-1}$ cm$^{-2}$)} & \colhead{Width (\AA)} }
\startdata
SSC 1 & 7.26(0.78) & 61.17(6.48) & 2300(410) \\ 
SSC 2 & 4.65(0.50) & 37.69(4.05) & 1870(330)  \\ 
SSC 3 & 0.55(0.07) & 6.18(0.78) & 580(110)  \\ 
SSC 4 & 0.23(0.04) & 1.55(0.56) & 80(30) \\ 
SSC 5 & 0.18(0.07) & 1.19(0.96) & 30(20)  \\ 
SSC 6 & 0.08(0.01) & 0.23(0.13) & 50(30) \\ 
\enddata
\end{deluxetable*}

\section{Properties of the SSCs}\label{propsec}

\subsection{Spectral Energy Distributions}\label{sedsec}

We estimate the physical properties of the SSCs in SBS 0335-052 by comparing
their spectral energy distributions (SEDs) to the latest STARBURST99
population synthesis models (v5.1) of \citet{Leitherer99}.  We have run two
simulations using different evolutionary tracks and metallicities: (1) the
Geneva tracks with high mass loss and Z=0.001 and (2) the Padova tracks with
AGB stars and Z=0.0004.  In both cases, we adopt an instantaneous burst of
$10^5 M_\odot$ with a Kroupa IMF \citep{Kroupa01} and the Pauldrach/Hillier atmospheres.
The two simulations use the lowest metallicity evolutionary tracks available 
in STARBURST99 from the Geneva and Padova groups, appropriate for SBS 0335-052.
The high mass loss tracks are recommended by the Geneva group and the inclusion
(or exclusion) of AGB stars in the Padova tracks is not critical since we
are modelling very young clusters.  We have run the two simulations for comparison
and as a consistency check. 

\begin{figure*}[!t]
\begin{center}$
\begin{array}{cc}
\includegraphics[width=2.65in]{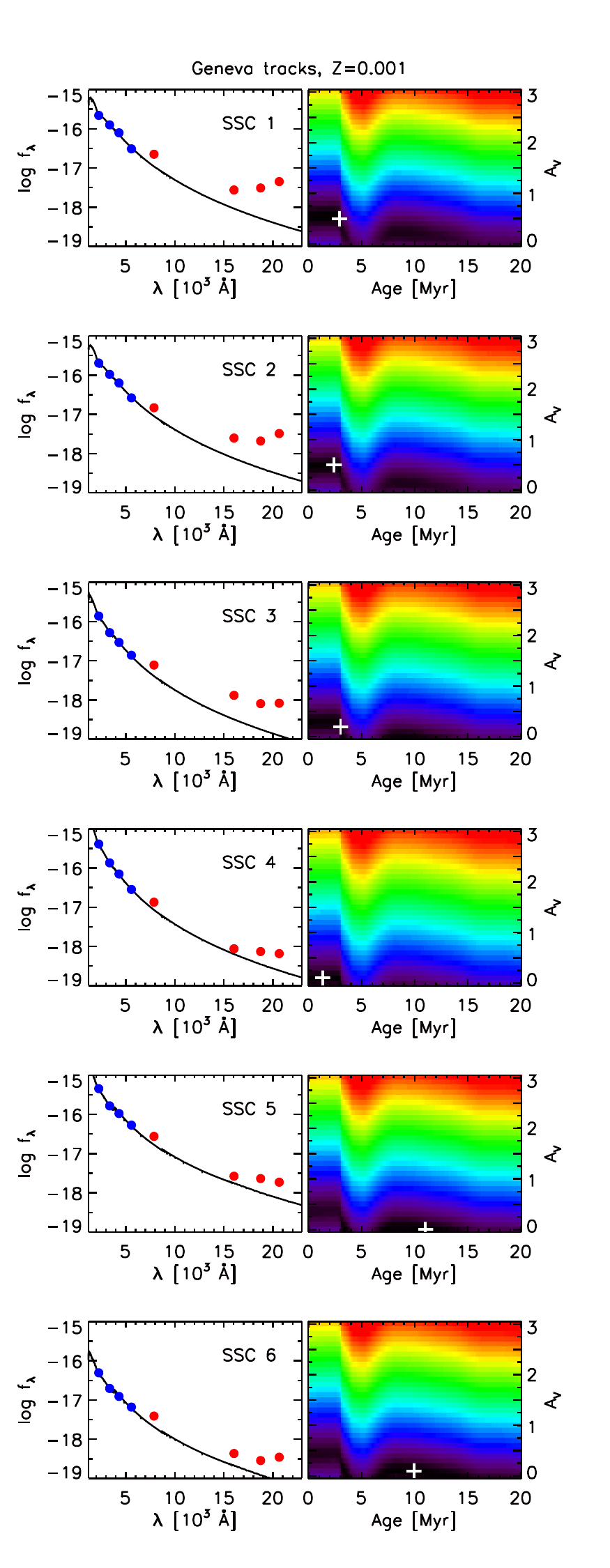} &
\includegraphics[width=2.65in]{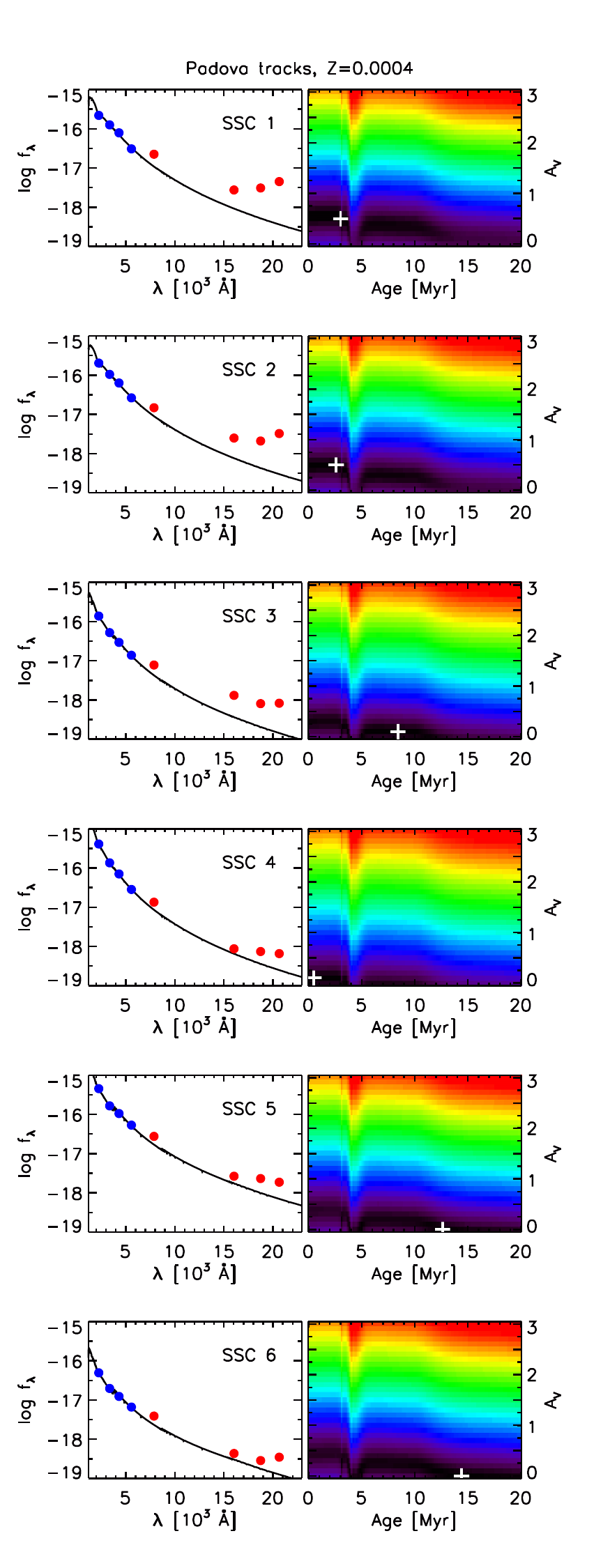}
\end{array}$
\caption{UV to near-IR observed flux densities (erg s$^{-1}$ cm$^{-2}$
\AA$^{-1}$) and best-fitting STARBURST99 model SEDs (using the Geneva (left) and Padova (right)
evolutionary tracks) for the SSCs in SBS 0335-052.  The blue data
points (F220W, F330W, F435W, F550M) are included in the fitting process but the red points
(F791W, F160W, F187N, F205W) are not.  Grids of $\sigma$ (the
standard deviation of the logarithmic residuals of the observed and model
flux densities) are shown for models of various ages and extinctions.
Black indicates better fitting models (lower values of $\sigma$) and red indicates worse
fitting models (higher values of $\sigma$).  A white cross indicates the best-fitting
model.  The dip in $A_V$ at $\sim 5$ Myr is the result of a dramatic reddening of
the intrinsic model colors (e.g. $U-B$) at this time.  Lower values of $A_V$
are preferred during this time since the intrinsic colors are already very red.
\label{sedplots}}
\end{center}
\end{figure*}

Ages, extinctions, and masses of the SSCs are estimated by comparing a grid
of model SEDs (separately for each simulation) to the broad-band {\it HST}
flux densities.  The grids consist of model SEDs with ages up to 20 Myr in
increments of $\Delta t=0.1$ Myr and extinctions up to $A_V=3$ mag in
increments of 0.1 mag.  The work of \citet{Reines08} on
young massive clusters in the dwarf starburst galaxy NGC 4449 showed that model
SEDs modified by a 30 Doradus extinction curve\footnote{The 30 Doradus
extinction curve is adopted from \citet[][Table 3]{Misselt99} and \citet[]
[Table 6]{Fitzpatrick85}, using the parameterization given by
\citet{Fitzpatrick90}.} fit the clusters' broadband flux densities
significantly better than the Galactic extinction curve \citep{Cardelli89}
and the starburst obscuration curve \citep{Calzetti00}.  Therefore, we first
apply the 30 Doradus extinction curve with a given $A_V$ to the model
SEDs.  Next, we simulate Galactic foreground extinction towards SBS 0335-052
\citep[$E(B-V)=0.047$ mag,][]{Schlegel98} using the extinction curve
of \citet{Cardelli89}.  Each model SED, of a given age and $A_V$, is
then convolved with the {\it HST} total throughput curves of the appropriate
filters before it is compared to the measured flux densities.
The best-fit model SED is determined by minimizing a goodness-of-fit parameter,
$\sigma$, equal to the standard deviation of the logarithmic residuals of the
observed and model flux densities.
A mass estimate is obtained by scaling the model mass by the mean logarithmic
offset between the observed and best-fit model flux densities.

We include the F220W, F330W, F435W, and F550M flux densities in the SED
fitting.  The strongest Balmer lines, H$\alpha$ and H$\beta$, are not
contained in this filter set, although the F435W filter does include
H$\gamma$ and H$\delta$.  Since \citet{Reines08} revealed excesses in
the $I$- and $H$-band flux densities of the young clusters in NGC 4449
compared to model SEDs, we did not include data at wavelengths longer
than the F550M filter ($\sim V$) in the fitting process here.  
The observed and best-fitting model SEDs of the SSCs in SBS 0335-052
are shown in Figure \ref{sedplots}.  The grids of $\sigma$ are also
shown to give a sense of the uncertainty in the estimates of ages
and extinctions.  Black indicates lower values of sigma (better fits) and
red indicates higher values of sigma (worse fits).  The best-fitting
model is indicated by a white cross-point.
Figure \ref{sedplots} also shows that there is a clear excess in the
$I$-band and at near-IR wavelengths compared to the model SEDs.
We will return to this in \S\ref{excess_sec}.

Estimates of the ages, masses, and extinctions of the SSCs
are found by averaging these physical properties of the best-fitting
model SEDs from the two STARBURST99 simulations and the results are listed
in Table \ref{sedprop}.  We note that the derived extinctions for SSCs 1 and
2, $A_V \sim 0.5$, are consistent with the extinction from the optical
recombination line ratio H$\alpha$/H$\beta$
found by \citet{Izotov97}.  There is no significant difference between the
goodness-of-fits using the two different STARBURST99 simulations.  The
quoted errors reflect the variations in the results using the two simulations,
although they represent the minimum uncertainties since the models are
somewhat degenerate in age and extinction (see Figure \ref{sedplots}). 
Despite this degeneracy, we still have confidence in the physical properties of the SSCs
derived from SED fitting for the following reasons.  First, the ages are
consistent with those we derive from H$\alpha$ equivalent widths in the
following section (with the exception of SSC 4, see below).  Also, as noted above, the
$A_V$ are in agreement with the work of \citet{Izotov97}, although
we emphasize that caution must be applied when interpreting the meaning of extinction
estimates of the embedded SSCs (1 and 2).  We will address this issue in detail in
\S\ref{ext_sec}. 

\begin{deluxetable}{cccc}
\tabletypesize{\footnotesize}
\tablecolumns{4}
\tablewidth{0pt}
\tablecaption{Properties from SED Fitting \label{sedprop}}
\tablehead{
\colhead{Source} & \colhead{Age} & \colhead{Mass} & \colhead{$\rm A_V$} \\
\colhead{ } & \colhead{(Myr)} & \colhead{($10^6 M_\odot$)} & \colhead{(mag)} }
\startdata
SSC 1 & $\lesssim 3$\tablenotemark{a} & 1.0(0.1) & 0.5(0.1) \\ 
SSC 2 & $\lesssim 3$\tablenotemark{a} & 1.1(0.1) & 0.5(0.1) \\ 
SSC 3 & 5.7(3.8) & 0.4(0.2) & 0.2(0.1) \\ 
SSC 4 & $\lesssim 3$\tablenotemark{a,}\tablenotemark{b} & 1.1(0.1) & 0.1(0.1) \\ 
SSC 5 & 11.8(1.1) & 1.8(0.2) & 0.0(0.1) \\ 
SSC 6 & 12.2(3.2) & 0.2(0.1) & 0.1(0.1) \\ 
\enddata
\tablecomments{Estimates of the physical properties of the SSCs in
SBS 0335-052 derived from SED fitting.  The values listed are
averages from the best-fitting model SEDs using the two STARBURST99
simulations described in the text (there was no significant difference
in the fits using either of the model simulations).  The quoted errors in
parenthesis reflect the variations in the results using the two simulations.
Also see Figure \ref{sedplots}.}
\tablenotetext{a}{The SEDs are insensitive to ages $\lesssim 3$ Myr.}
\tablenotetext{b}{This young age is probably incorrect since copious
amounts of H$\alpha$ and P$\alpha$ emission would be expected for such
a young cluster and this is not observed.  The age of SSC 4 inferred from
the H$\alpha$ equivalent width is $\sim 12$~Myr.}
\end{deluxetable}

\subsection{Nebular Emission from the SSCs}

\subsubsection{Ages}\label{agesec}

In addition to SED fitting to the stellar continuum, the nebular emission from
a starburst region can also provide estimates of the physical properties of the
host cluster.  Age estimates can be obtained from the equivalent
width of H$\alpha$ emission since this quantity measures the ratio
of ionizing flux (from massive, short-lived stars) to the total
continuum flux density.  This ratio is strongly dependent on age for
stellar populations between approximately 3 and 20 Myr old.  The insensitivity
before $\sim 3$ Myr arises because the most massive stars must
begin to die in order for the ionizing flux (and H$\alpha$ equivalent width) to
decrease.  Clusters older than $\sim 20$ Myr are not strong H$\alpha$ emitters.

In addition, if the dusty interstellar medium (ISM) surrounding the clusters
is uniform and every Lyman continuum
photon ionizes a hydrogen atom, rather than getting absorbed by dust or escaping
the cluster altogether, then
H$\alpha$ equivalent width will be independent
of reddening (since the continuum is measured at the wavelength of H$\alpha$).
If, however, the surrounding ISM is porous and a significant
fraction of ionizing and non-ionizing photons escape the cluster,
the equivalent width of H$\alpha$ will be artificially low and only provide
an upper limit on the age of the cluster.  This is because we would
detect all of the stellar continuum photons at 6563 \AA~that escape, but
the H$\alpha$ flux would be reduced due to the leakage of ionizing photons
(which do not actually ionize the gas).

Age estimates from comparing the H$\alpha$ equivalent widths of the SSCs
with the STARBURST99 models are given in Table \ref{lineprop} and a significant
age spread between the clusters is found ($\lesssim 3$ to $\sim 15$ Myr).  These age
estimates are consistent with those obtained from SED fitting (within the errors), with the exception
of SSC 4.  In this case, the age from SED fitting ($\lesssim 3$ Myr) is most likely incorrect 
since we would expect copious amounts of H$\alpha$ and Pa$\alpha$ emission
from such a young cluster and this is not observed.  The age of SSC 4 from the H$\alpha$ equivalent
width is $\sim 12$ Myr.  It is not entirely clear why such a young age was found from SED
fitting, although contamination by the diffuse emission surrounding SSC 4 (see Figure \ref{allims})
at short wavelengths may be responsible.

\subsubsection{Ionizing Luminosities}\label{ion}

The production rate of Lyman continuum photons, $Q_{\rm Lyc}$, from the SSCs
can be estimated from their H$\alpha$ and Pa$\alpha$ fluxes.  From Equations 2 and
3 in \citet{Condon92}, the ionizing photon rate as a function of H$\beta$ luminosity,
$L_{{\rm H}\beta}$, is given by

\begin{equation}
\left({Q_{\rm Lyc} \over {\rm s^{-1}}}\right) \gtrsim 2.25\times10^{12}
\left({T_e \over 10^4{\rm ~K}}\right)^{0.07} \left({L_{{\rm H}\beta} \over {\rm erg ~s^{-1}}}\right).
\label{Qlyc}
\end{equation}

\noindent
This equation provides only a lower limit since a fraction of the ionizing photons
may be absorbed by dust, or escape the cluster altogether 
before ionizing hydrogen atoms if the ISM is porous and clumpy.
For Case B recombination, an electron temperature of 20,000 K \citep{Izotov97}, and
an electron density between $10^2$ and $10^4~{\rm cm}^{-3}$, the expected
(reddening-free) flux ratios of H$\alpha$ and Pa$\alpha$ to H$\beta$ are given by
\citet{Condon92} and \citet{Osterbrock89}: $F_{{\rm H}\alpha}/F_{{\rm H}\beta}=2.73$,
$F_{{\rm Pa}\alpha}/F_{{\rm H}\beta}=0.28$.  We use Equation \ref{Qlyc} and the predicted
line ratios with our observed (not extinction corrected) H$\alpha$ and Pa$\alpha$
fluxes to obtain estimates of the ionizing luminosities, $Q_{\rm Lyc}$, produced by
the SSCs in SBS 0335-052 (see Table \ref{lineprop}).  

In addition to the H$\alpha$ and Pa$\alpha$ fluxes, we use new measurements of the
(thermal) free-free radio flux densities at 1.3 cm \citep{Johnson08} to estimate the
ionizing luminosities of SSCs 1 and 2 (the other clusters do not have significant
detections at 1.3 cm).  \citet{Johnson08} find that the radio flux densities at 1.3 cm
are optically thin and equal to 0.17 mJy and 0.12 mJy for SSCs 1 and 2, respectively.
We adopt a distance of 54 Mpc to calculate the radio luminosity densities,
$L_{\nu, \rm thermal}$, and follow \citet{Condon92}

\begin{eqnarray}
\left({Q_{\rm Lyc} \over {\rm s^{-1}}}\right) \gtrsim 6.3\times10^{52}
\left({T_e \over 10^4{\rm ~K}}\right)^{-0.45} \left({\nu \over {\rm GHz}}\right)^{0.1} \\
\times \left({L_{\nu, \rm thermal} \over 10^{27} {\rm ~erg ~s^{-1} ~Hz^{-1}}}\right).
\nonumber
\end{eqnarray}

\noindent
to estimate the ionizing luminosities of SSCs 1 and 2 (see Table \ref{lineprop}).

\begin{figure}
\begin{center}
\includegraphics[scale=.66]{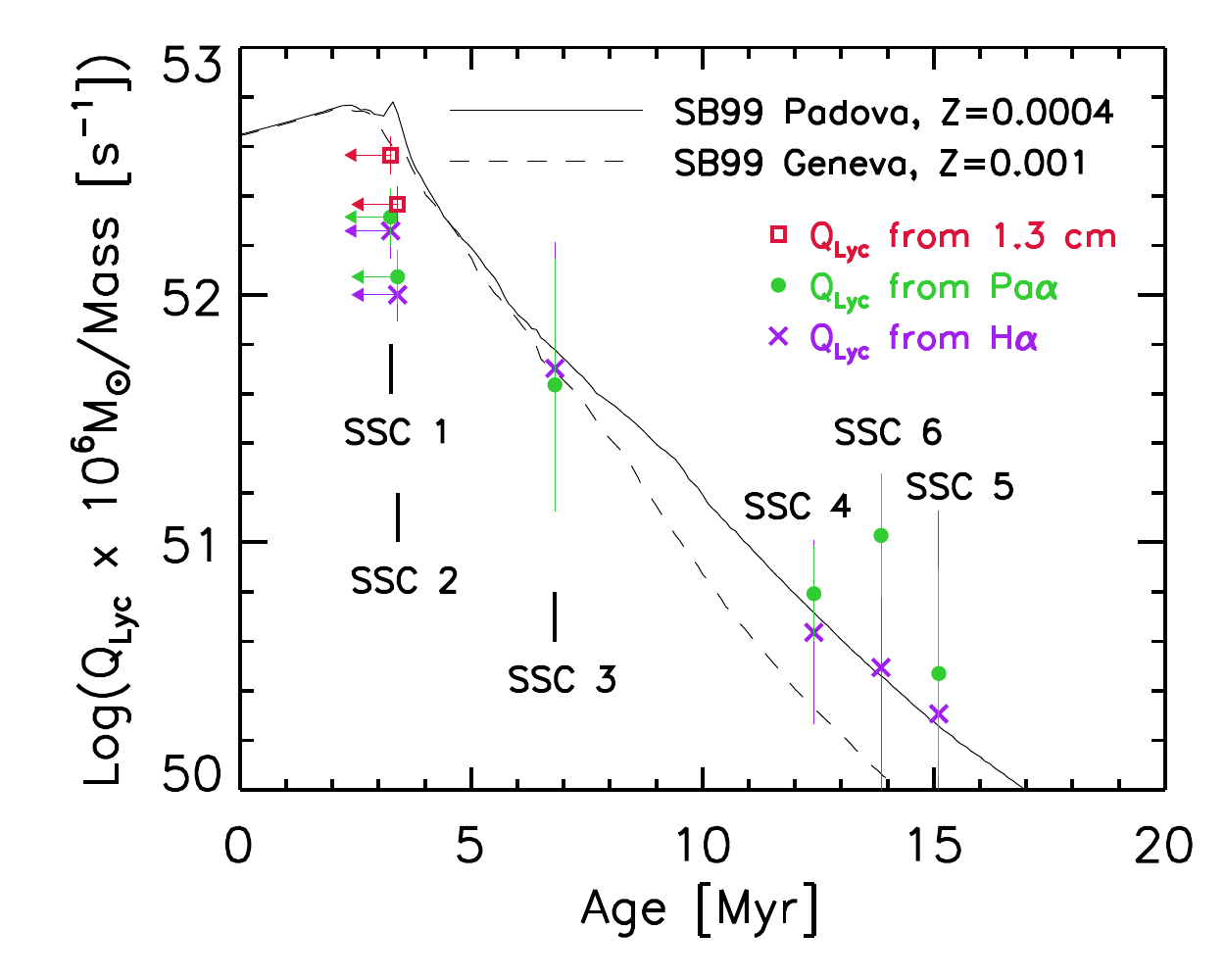}
\caption{Scaled ionizing fluxes versus age (from H$\alpha$ equivalent width).
For SSCs 1 and 2, the ionized gas emission underpredicts the expected values
from the non-ionizing stellar continuum.  This suggests the ISM is porous
and stellar continuum photons (both ionizing and non-ionizing) are leaking
out of the immediate vicinity of the clusters.  See the text for a full
discussion.
\label{nly_age}}
\end{center}
\end{figure}

\begin{deluxetable*}{lccccccc}
\tabletypesize{\footnotesize}
\tablecolumns{8}
\tablewidth{0pt}
\tablecaption{Properties from the Nebular Emission \label{lineprop}}
\tablehead{
\colhead{ } & \colhead{Age} & \colhead{$Q_{\rm Lyc}^{{\rm 1.3 cm}}$} & 
\colhead{$Q_{\rm Lyc}^{{\rm Pa}\alpha}$} & \colhead{$Q_{\rm Lyc}^{{\rm H}\alpha}$} &
\colhead{$f_{Q}^{{\rm 1.3 cm}}$} & \colhead{$f_{Q}^{{\rm Pa}\alpha}$} & \colhead{$f_{Q}^{{\rm H}\alpha}$} \\
\colhead{ } & \colhead{(Myr)} & \colhead{($10^{49}$~s$^{-1}$)} &
\colhead{($10^{49}$~s$^{-1}$)} & \colhead{($10^{49}$~s$^{-1}$)} & \colhead{ } & \colhead{ } & \colhead{ } }
\startdata
SSC 1\tablenotemark{*} & $\lesssim 3.3$ & $\gtrsim 3750$ & $\gtrsim 2110$ & $\gtrsim 1860$ & 0.76(0.16) 
& 0.43(0.09) & 0.38(0.08) \\
SSC 2\tablenotemark{*} & $\lesssim 3.4$ & $\gtrsim 2650$ & $\gtrsim 1350$ & $\gtrsim 1140$ & 0.51(0.11)
& 0.26(0.06) & 0.22(0.05) \\
SSC 3 & 6.8(2.5) & \nodata & 160(20) & 190(20) & \nodata & $\sim 1$ & $\sim 1$ \\
SSC 4 & 12.4(1.7) & \nodata & 70(10) & 50(20) & \nodata & $\sim 1$ & $\sim 1$ \\
SSC 5 & 15.1(2.3) & \nodata & 50(20) & 40(30) & \nodata & $\sim 1$ & $\sim 1$ \\
SSC 6 & 13.9(1.9) & \nodata & 20(10) & 10(10) & \nodata & $\sim 1$ & $\sim 1$ \\
\enddata
\tablecomments{Age estimates of the SSCs from H$\alpha$ equivalent widths, as well as
measured ionizing fluxes ($Q_{\rm Lyc}$) and recovered fractions of ionizing fluxes ($f_Q$)
from H$\alpha$, Pa$\alpha$ and 1.3 cm radio flux densities \citep{Johnson08}.  A three-dot
ellipsis indicates sources that do not have detections at 1.3 cm.}
\tablenotetext{*}{The age estimates for these young clusters are upper limits since
H$\alpha$ equivalent width is insensitive to ages $\lesssim 3$ Myr and also
possibly due to the leakage of ionizing photons through a porous ISM (see discussion
in the text).  The ionizing fluxes for SSCs 1 and 2 are lower limits
because they do not reflect ionizing photons that either escape through a porous
ISM or are absorbed by dust. Correcting the $Q_{\rm Lyc}$ by the $f_Q$'s imply SSCs
1 and 2 each have ionizing fluxes of $\sim 5 \times 10^{52}~{\rm s}^{-1}$.} 
\end{deluxetable*}

The ionizing luminosities of the SSCs should decrease as they age.  Since we have
established an age spread between the different clusters, we can check this
prediction.  However, ionizing luminosity is also a function of mass (in addition
to age) so we need to scale the measured $Q_{\rm Lyc}$ values to a common mass and remove
the mass dependence.  To achieve this, we divide the measured $Q_{\rm Lyc}$ values by the
masses derived from SED fitting (Table \ref{sedprop}) and multiply by $10^6~M_\odot$ to
scale the ionizing luminosities to those expected for a $10^6~M_\odot$ cluster.  Figure
\ref{nly_age} shows the scaled ionizing luminosities versus cluster age
(derived from H$\alpha$ equivalent widths).  The two STARBURST99 models are also shown
for comparison.  

From Figure \ref{nly_age}, we see that the ionizing luminosities of SSCs 3-6, derived
from their H$\alpha$ and Pa$\alpha$ fluxes, are consistent with the model predictions.
However, the ionizing luminosities of SSCs 1 and 2 (ages $\lesssim$ 3 Myr), derived from
their H$\alpha$ and Pa$\alpha$ fluxes, are well below
the expected values.  In other words, these recombination lines are only sampling a
fraction of the ionizing photons relative to the number expected from the non-ionizing  
stellar continuum inferred from the UV and optical photometry (recall that we have
normalized the $Q_{\rm Lyc}$ values by the masses derived from SED fitting to the UV
and optical photometry).  Equivalently, the masses inferred from the recombination line
fluxes are less than those derived from SED fitting.

Interestingly, the optically-thin, free-free radio emission
from SSCs 1 and 2 \citep{Johnson08} also underpredicts the
ionizing luminosities (see Figure \ref{nly_age}).
This may suggest the ISM surrounding the young SSCs is clumpy and porous,
with virtually unobscured lines-of-sight to the cluster stars.  Clearly, some fraction
of stellar continuum photons escape the clusters through low-extinction regions: we
observe copious amounts of UV and optical photons and the $A_V$ derived from
SED fitting is a modest $\sim 0.5$ mag for SSCs 1 and 2.  
The lower than expected $Q_{\rm Lyc}$ suggest ionizing photons also escape
the immediate vicinity of the clusters (without ionizing hydrogen), resulting in the
loss of H$\alpha$, Pa$\alpha$ and radio emission.  This in turn, implies
the H$\alpha$ equivalent widths are lower limits and the ages of SSCs 1 and 2 are
$\lesssim 3$ Myr.

We can calculate the recovered fractions of the ionizing luminosities from SSCs 1 and 2
by comparing the data points to the models in Figure \ref{nly_age}.  
The fraction of ionizing luminosity recovered
by H$\alpha$, $f_Q^{{\rm H}\alpha}$, is $\sim 0.38$ for SSC 1 and $\sim 0.22$ for SSC 2.
These values are consistent with previous estimates that, in the optical, only about
1/4 of the total ionized gas emission is observed for the region containing SSCs 1 and
2 \citep{Hunt01}.  From Pa$\alpha$, $f_Q^{{\rm Pa}\alpha} = 0.43~({\rm SSC~1)~and}~0.26~({\rm SSC~2})$.
Using the radio-derived ionizing luminosities, we find $f_Q^{{\rm 1.3 cm}}$ equals $\sim 0.76$
and $\sim 0.51$ for SSCs 1 and 2, respectively.  The increase of $f_Q$ with wavelength
naturally follows from differential absorption (i.e. reddening) of the ionized gas as probed
by the H$\alpha$, Pa$\alpha$ and radio emission.

The measured ionizing luminosities and the recovered fractions of the intrinsic
ionizing luminosities are summarized in Table \ref{lineprop}.
Correcting the measured ionizing luminosities (i.e. dividing the $Q_{\rm Lyc}$'s by $f_Q$'s)
implies the intrinsic ionizing luminosities of SSCs 1 and 2 are each
$\sim 5 \times 10^{52}~{\rm s}^{-1}$, or the equivalent of $\sim 5000$ O7.5 V
stars \citep{Vacca96}.  The inferred masses of these SSCs are $\sim 10^6 M_\odot$
(see Table \ref{sedprop}).

\subsubsection{Extinctions of the Embedded SSCs}\label{ext_sec}

The youngest SSCs in SBS 0335-052 (1 and 2) are still embedded in their
natal birth material: they are the strongest emitters of H$\alpha$ and Pa$\alpha$
and the only clusters in the galaxy detected in the radio via free-free emission \citep{Johnson08}.   
Various studies have provided extinction estimates from recombination lines
of the region encompassing these two clusters (previous studies were not able to resolve
the two sources) and the derived values {\it appear} to be discrepant:
$A_V=0.55$ from H$\alpha$/H$\beta$ \citep{Izotov97}, $A_V=0.73$ from H$\beta$/Br$\gamma$
\citep{Vanzi00}, $A_V=1.45$ from H$\beta$/Br$\alpha$ \citep{Hunt01}, and
$A_V=12.1$ from Br$\gamma$/Br$\alpha$ \citep{Hunt01}.  Mid-IR observations complicate
the situation further.  \citet{Thuan99} found that the mid-IR SED is well-described
by a modified blackbody spectrum extinguished by a screen of dust with $A_V \sim 20$ mag.
A qualitatively similar result was found by \citet{Plante02} and \citet{Hunt05}.
These authors model the observed mid-IR SED using DUSTY \citep{Ivezic97,Ivezic99}, a
program that solves the radiation transfer equations in a spherical
environment, and conclude that the embedded starburst suffers $\sim 30$ magnitudes
of visual extinction.  \citet{Houck04} also find a large extinction, $A_V \sim 15$, 
from the 9.7 $\mu$m silicate absorption feature assuming a screen model and a blackbody
background source.  In this section, we attempt to reconcile the apparently discrepant
extinction estimates assuming all of the emission (at all wavelengths) is associated with
the optically visible SSCs 1 and 2.

We begin by estimating the extinctions of the recombination lines H$\beta$ \citep{Izotov97},
H$\alpha$ (this work), Pa$\alpha$ \citep[][and this work]{Thompson06}, Br$\gamma$
\citep{Vanzi00}, and Br$\alpha$ \citep{Hunt01} relative to the (thermal) free-free radio
emission from this region \citep{Johnson08}.  The radio
emission does not suffer from extinction {\it per se}, although again we note that the
far-UV photons can be absorbed by dust or escape the boundaries of the clusters before
ionizing the gas.  In either of these scenarios, the ionizing flux probed by the
optical and IR lines would also be reduced by the same amount and the relative extinction
between these lines and the radio would not be affected.  To obtain the extinction
estimates, we compare the measured fluxes to the expected values predicted by the
free-free radio emission at 1.3 cm.  Following \citet{Condon92}, the predicted
H$\alpha$ flux is given by

\begin{eqnarray}
\left({F_{{\rm H}\alpha,{\rm predicted}} \over {\rm erg~s^{-1}~cm^{-2}}}\right) \sim 0.8 \times 10^{-12}
\left({T_e \over 10^4{\rm ~K}}\right)^{-0.59} \\
\times \left({\nu \over {\rm GHz}}\right)^{0.1} \left({S_{\nu, \rm thermal} \over {\rm mJy}}\right).
\nonumber
\end{eqnarray}

\noindent
where $S_{\nu, \rm thermal}$ is the thermal free-free radio flux density (= 0.56 mJy).    
To obtain the predicted fluxes for the other recombination lines, we use the
theoretical (reddening-free) line ratios appropriate for SBS 0335-052
(Case B, T=20000 K) given by \citet{Condon92} and \citet{Osterbrock89}.
The observed recombination line fluxes and the radio-predicted values are listed
in Table \ref{Hlines}.  We give the flux ratios (observed to predicted) and the
corresponding extinctions of these lines at their native wavelengths, $A_\lambda$.

\begin{deluxetable}{cccccc}
\tabletypesize{\footnotesize}
\tablecolumns{6}
\tablewidth{0pt}
\tablecaption{Recombination Lines from the Region SSC 1+SSC 2 \label{Hlines}}
\tablehead{
\colhead{Line} & \colhead{Rest $\lambda$} & \colhead{Observed} & \colhead{Predicted} & 
\colhead{Flux} & \colhead{$A_\lambda$} \\
\colhead{ } & \colhead{($\mu$m)} & \colhead{Flux\tablenotemark{a}} & 
\colhead{Flux\tablenotemark{a}} & \colhead{Ratio} &\colhead{(mag)} }
\startdata
H$\beta$   & 0.486 & 6.1  & 14.3 & 0.4 & 0.9  \\
H$\alpha$  & 0.656 & 16.6 & 40.8 & 0.4 & 1.0 \\
Pa$\alpha$ & 1.875 & 1.77 & 4.05 & 0.4 & 0.9 \\
Br$\gamma$ & 2.166 & 0.15 & 0.34 & 0.4 & 0.9 \\
Br$\alpha$ & 4.052 & 0.90 & 0.88 & 1.0 & 0.0 \\
\enddata
\tablecomments{Observed recombination line fluxes for H$\beta$ \citep{Izotov97},
H$\alpha$ (this work), Pa$\alpha$ \citep[][and this work]{Thompson06}, Br$\gamma$
\citep{Vanzi00}, and Br$\alpha$ \citep{Hunt01} were measured
in a $\sim 1\farcs5 \times 1\farcs0$ aperture.  The predicted values are from the
free-free radio flux density at 1.3 cm \citep{Johnson08}.}
\tablenotetext{a}{Units are $10^{-14}$ erg s$^{-1}$ cm$^{-2}$.}
\end{deluxetable}

It is quite intriguing that, relative to the free-free radio emission, the calculated
extinctions of the H$\beta$, H$\alpha$, Pa$\alpha$, and Br$\gamma$ emission are all
$\sim 1$ mag (at their native wavelengths between $\sim 0.5-2~\mu$m).  The similar
estimates of $A_\lambda$ (or equivalently, the observed to predicted flux ratios)
for these lines appear to indicate that there is not a significant amount of
{\it reddening} (i.e. differential absorption with short wavelengths suffering
more absorption) of the ionized gas between $\sim 0.5-2~\mu$m.

The value $A_\lambda \sim 1$ then gives the total amount of {\it absorption}
of these lines compared to the radio (i.e. the ratio of the ionized gas probed by
the optical/near-IR lines to the total amount of ionized gas inferred from the
free-free radio emission).
If the ISM is clumpy as we have previously suggested (\S\ref{ion}) and no optical
or near-IR light gets through dense dust clumps (i.e. the light is completely
absorbed), the ratio of the optical (and near-IR)
emission to the radio emission is equivalent to the ratio of the amount of flux
that emerges through the ``holes'' (optical/near-IR) to the amount of flux that emerges
through both the holes and the clumps (radio).  This ratio equals 0.4 for 
H$\beta$, H$\alpha$, Pa$\alpha$, and Br$\gamma$.  In other words, $\sim 40\%$ of the
dust cocoon is void of clumps (low extinction regions) and $\sim 60\%$ is covered by
clumps of dust that totally absorb light shortward of $\sim 2~\mu$m (high extinction
regions).  In effect, there is a clumpy dust covering factor of $\sim 60\%$.

These values are strikingly similar to our estimates of the fraction
of ionizing luminosity recovered by the radio emission ($\sim 64\%$ on average for
SSCs 1 and 2) and the fraction of ionizing luminosity that escapes the clusters
undetected ($\sim 37\%$, see \S\ref{ion}).  This {\it may} suggest that the dust clumps
and the neutral gas clumps are coincident, although this is not necessarily the case.

\begin{figure}
\begin{center}
\includegraphics[scale=0.33]{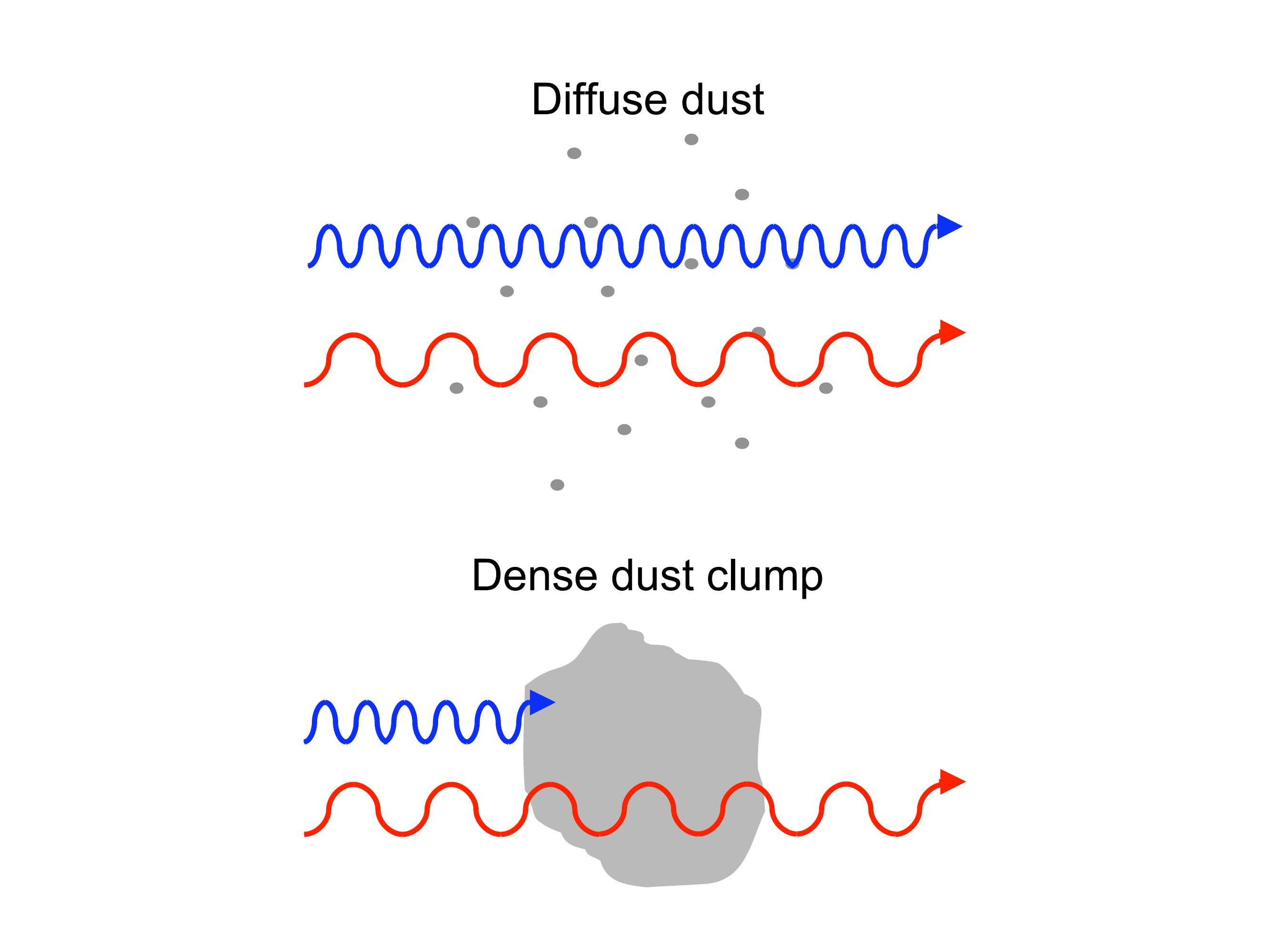}
\caption{A cartoon illustrating the effects of a clumpy dust structure surrounding
the young SSCs (1 and 2) in SBS 0335-052.  Short-wavelength photons (blue) with
$\lambda \lesssim 2~\mu$m (i.e. stellar continuum, H$\beta$, H$\alpha$, Pa$\alpha$,
and Br$\gamma$) are completely absorbed by dense dust clumps harbouring large grains
and only emerge through the diffuse interclump regions, accounting for the low {\it measured}
extinctions in the optical/near-IR.  Longer-wavelength
photons (red) with $\lambda \gtrsim 4~\mu$m (i.e. Br$\alpha$ and 1.3 cm) can penetrate the
dense clumps as well as the diffuse regions.  See \S\ref{ext_sec} for a full discussion.}
\label{dustcartoon}
\end{center}
\end{figure}

The picture outlined above is illustrated schematically in Figure \ref{dustcartoon}.
The large extinctions ($A_V \gtrsim 15-30$) derived from mid-IR observations
\citep{Thuan99,Plante02,Houck04,Hunt05} would be a natural consequence {\it if} the
mid-IR emission originates from the dense dust clumps and not a uniform screen.  
In our proposed scenario, the dust would be heated by the strong UV stellar continuum
impinging on the clumps.  This would help explain the large extinction derived from the
mid-IR spectral energy distributions.  Also, if the extinctions inferred from the mid-IR
describe absorption from dust clumps rather than a screen of dust, they may actually
underestimate the absorption because of beam dilution effects.  In addition, diffuse
dust in the interclump regions could provide the modest reddening measured
at optical wavelengths (since the optical light only emerges through the low
extinction regions).  The scenario described here is similar to the one proposed by
\citet{Gordon97} in which a dust-free star cluster is surrounded
by a clumpy shell of dust and gas.

Moreover, the extinction curve in SBS 0335-052 appears to be anomalous.  It is roughly
constant up to $2~\mu$m with $A_\lambda\sim 1$, then drops to $A_\lambda\sim 0$ at $4~\mu$m.
This behavior is qualitatively similar to extinction curves along lines of sight with high
gas densities.  Such regions have typically flatter extinction curves, thought to be due to
larger dust grains than in the diffuse ISM \citep[e.g.,][]{Kim94}.
In the case of SBS 0335-052, the grains may be as large as
$\sim 1-2~\mu$m; however the silicate absorption feature at $9.7~\mu$m 
\citep[e.g.,][]{Houck04} makes grains larger than this rather unlikely,
because in that case the absorption feature would be absent \citep{Maiolino02}.

\subsection{$I$-band and Near-IR Excesses}\label{excess_sec}

\begin{deluxetable}{ccccc}
\tabletypesize{\footnotesize}
\tablecolumns{5}
\tablewidth{0pt}
\tablecaption{Near-IR Excesses of the SSCs \label{excess}}
\tablehead{
\colhead{Source} & \colhead{F791W} & \colhead{F160W} & 
\colhead{F187N} & \colhead{F205W} }
\startdata
SSC 1 &  0.7 &  1.2 &  2.0 & 
2.7 \\
SSC 2 &  0.5 &  1.3 &  1.8 & 
2.6 \\
SSC 3 &  0.6 &  1.5 &  1.6 & 
2.0 \\
SSC 4 &  0.5 &  0.3 &  0.8 & 
1.1 \\
SSC 5 &  0.5 &  0.4 &  0.9 & 
1.1 \\
SSC 6 &  0.6 &  0.6 &  0.8 & 
1.3 \\
\enddata
\tablecomments{Near-IR excesses (in magnitudes) of the SSCs with respect
to the best-fitting model SEDs (to the UV and optical data).  An excess is defined as 
2.5$\times$log($f_{\rm observed}/f_{\rm model}$), where
$f_{\rm observed}$ and $f_{\rm model}$ are the observed and model
flux densities, respectively.  The values listed are averages from
the best-fitting model SEDs using the two STARBURST99 simulations
described in the text (there was no significant difference in the
fits using either of the model simulations).}
\end{deluxetable}

The SSCs in SBS 0335-052 exhibit excesses in the $I$-band and at
near-IR wavelengths with respect to model SEDs (Figure \ref{sedplots}).
Excesses in the F791W ($\sim I$), F160W ($\sim H$), F187N ($\sim 1.9~\mu$m
continuum), and F205W ($\sim K$) filters are given in Table \ref{excess},
where an excess in magnitudes is defined as 2.5$\times$log($f_{\rm observed}/f_{\rm model}$)
($f_{\rm observed}$ and $f_{\rm model}$ are the observed and best-fitting model 
flux densities (\S\ref{sedsec}), respectively).  In this section, we investigate
the nature of these excesses.  We emphasize that the $I$-band ($\sim 0.8~\mu$m)
and near-IR ($\sim 1.6-2.1~\mu$m) excesses likely have different origins. 

\subsubsection{Excess in the $I$-band: Extended Red Emission?}

The $I$-band excess observed in the clusters of SBS 0335-052
ranges from $\sim$0.5--0.7 mag, similar to the excess
found in the young massive clusters of NGC 4449\footnote{We note
that the $I$-band observations of SBS 0335-052 and NGC 4449 come
from different instruments and slightly different filters (WFPC2/F791W
for this work and ACS/F814W for NGC 4449).}
\citep[$\sim 0.5$ mag on average;][]{Reines08}.  The primary
origin of the $I$-band excess in these two systems is quite possibly
the same.  An extensive discussion of possible origins of the $I$-band
excess observed in the young clusters of NGC 4449 is given in
\citet{Reines08} and the favored hypothesis is Extended Red
Emission (ERE) \citep[for a review of ERE, see][]{Witt04}.
ERE is a photoluminescent process in which an interstellar particle
absorbs a UV photon and then ultimately emits a photon in the
$\sim 6000-9000$ \AA~wavelength range.  ERE is typically observed as
a broad ($\sim 1000$ \AA) emission feature and it has been
detected in many astrophysical environments where both dust and
UV photons are present, including the Orion Nebula \citep{Perrin92}
and the 30 Doradus Nebula \citep{Darbon98}.  ERE provides a favorable
hypothesis for the $I$-band ($\sim 0.8~\mu$m) excess observed in the SSCs
of SBS 0335-052, but it cannot explain the excesses seen at near-IR
wavelengths ($\sim 1.6-2.1~\mu$m).

\subsubsection{Excess in the Near-IR: Hot Dust in the Younger SSCs and
Red Supergiants in the Older SSCs}

\begin{figure}
\begin{center}
\includegraphics[scale=0.65]{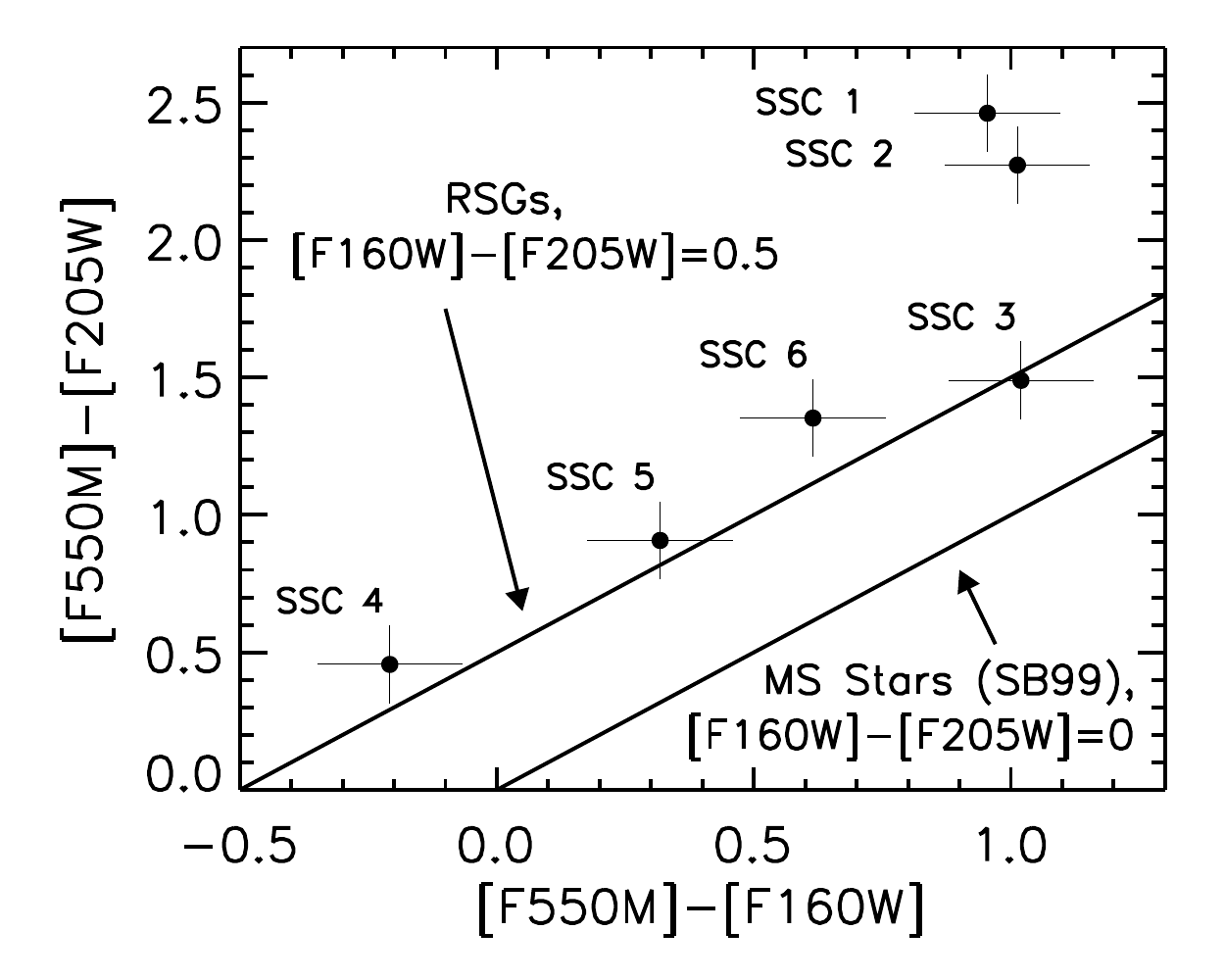}
\caption{[F550M]$-$[F205W] ($\sim V-K$) versus [F550M]$-$[F160W] ($\sim V-H$) color-color diagram.
The F550M magnitudes are dominated by the young hot stars in the clusters,
while the F160W and F205W magnitudes of all of the SSCs are much brighter
than predicted from the STARBURST99 models (Table \ref{excess}).
SSCs 3-6 (ages $\gtrsim 7$ Myr) have [F160W]$-$[F205W] colors consistent with red supergiants,
while SSCs 1 and 2 (ages $\lesssim 3$ Myr) have much redder colors
indicating another origin (see Figure \ref{IRseds}).
\label{vk_vh}}
\end{center}
\end{figure}

A near-IR/optical color-color diagram can provide insight into the origin
of the near-IR ($\sim 1.6-2.1~\mu$m) excesses observed in the SSCs of SBS 0335-052.
Figure \ref{vk_vh} shows a plot of [F550M]$-$[F205W] ($\sim V-K$) versus
[F550M]$-$[F160W] ($\sim V-H$) with the SSCs labeled.  The line [F160W]$-$[F205W]=0 is also shown,
approximating the color of main sequence stars throughout their lives and consistent with
the STARBURST99 evolutionary models.  In addition, we plot the line [F160W]$-$[F205W]=0.5, a characteristic
red supergiant (RSG) color \citep{McCrady03, Pickles98}\footnote{\citet{McCrady03}
calculate a synthetic [F160W]$-$[F222M] color of 0.52 for a M2 I star in the
\citet{Pickles98} library.}.  Note that in this color-color diagram, the
F550M magnitudes are dominated by the hot, unevolved cluster stars.

Figure \ref{vk_vh} illustrates that SSCs 3-6 have [F160W]$-$[F205W] colors consistent with RSGs.
This is in agreement with other studies showing these evolved stars are the dominant
source of near-IR light in SSCs older than $\sim 7$ Myr \citep[e.g.][]{McCrady03}.
We should also point out that RSGs could also contribute to the observed $I$-band
excess in these older clusters.
The two youngest SSCs in SBS 0335-052 (1 and 2) have extremely red [F160W]$-$[F205W] colors
that are quantitatively different from the other clusters: $\sim 1.5$ and
$\sim 1.3$, respectively (Table \ref{SSCphot}).  The origin of these
red colors is not consistent with RSGs and another mechanism must be responsible.
This is not surprising since RSGs should not have had enough time to evolve in
these $\lesssim 3$ Myr old clusters.  

Emission lines have been observed in the $H$ and $K$ spectra of SBS 0335-052
\citep{Vanzi00} in the region containing SSCs 1 and 2, however the line
contribution to the broad-band photometry here is negligible.
In addition, the near-IR excesses observed in SSCs 1 and 2 
(Figure \ref{sedplots}) are so large ($\sim 1.3$ mag in F160W,
$\sim 1.9$ mag in F187N, and $\sim 2.7$ mag in F205W; Table \ref{excess}),
and an excess is present in the narrow-band continuum filter F187N, that
continuum emission must be the dominant cause.  

\begin{figure}
\begin{center}$
\begin{array}{cc}
\includegraphics[width=3.5in]{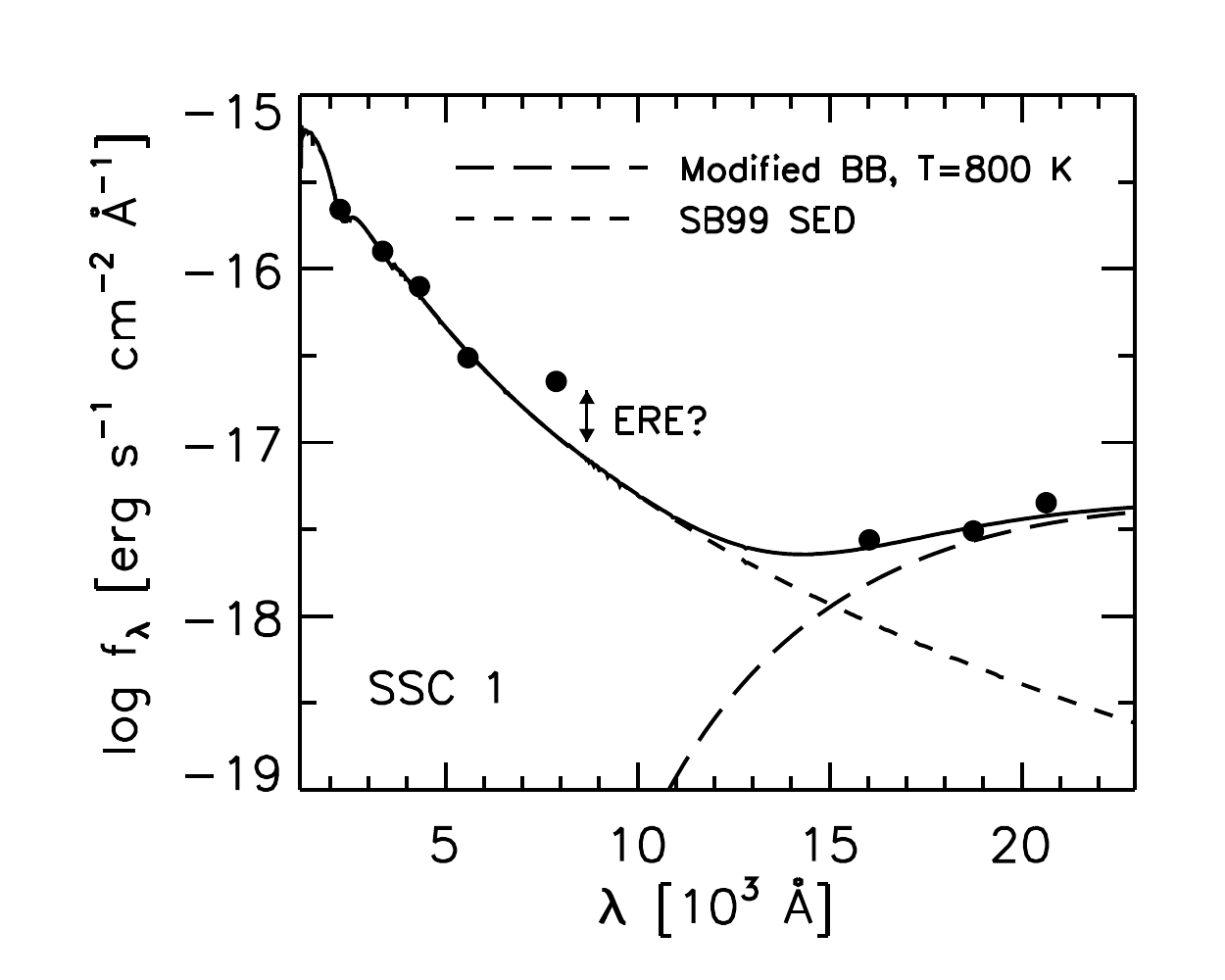} \\
\includegraphics[width=3.5in]{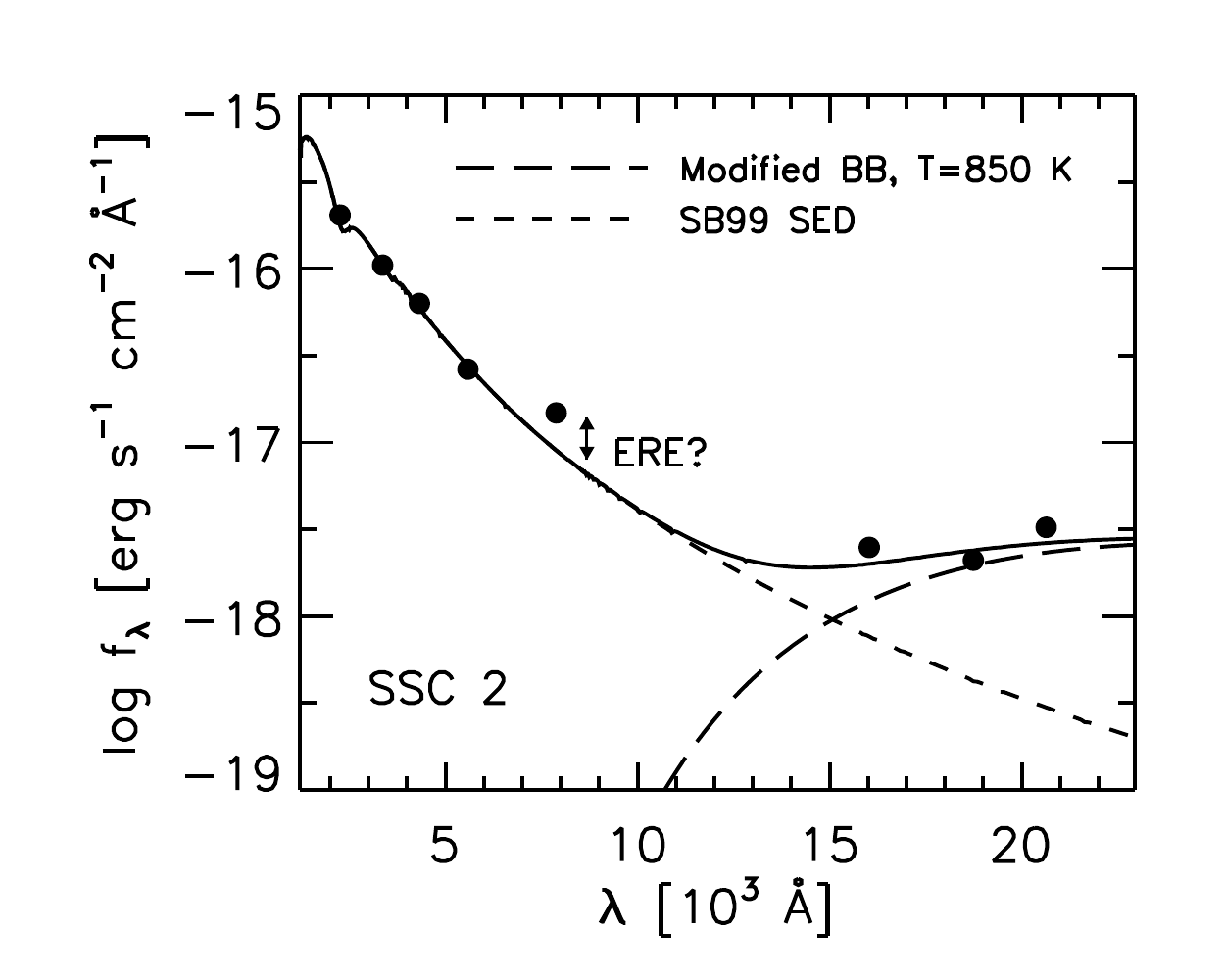}
\end{array}$
\caption{UV to near-IR spectral energy distributions for SSCs 1 and 2.
For each cluster, we add a hot dust component (modified blackbody) to
the best-fitting STARBURST99 model (to the UV and optical data) found in
\S\ref{sedsec}.  Temperatures equal to 800 K (SSC 1) and 850 K (SSC 2)
are effective at reproducing the near-IR excesses. 
\label{IRseds}}
\end{center}
\end{figure}

Although ionized gas continuum emission may be important
\citep[see e.g.][]{Vanzi00, Hunt01}, we propose that hot dust is the
main origin of the red [F160W]$-$[F205W] colors for SSCs 1 and 2.  To investigate
this hypothesis further, we add a modified blackbody with a dust opacity curve
given by \citet{Whitney03} and \citet{Kim94} to the best-fitting model SEDs
(to the UV and optical data) presented in \S\ref{sedsec}.  The results are positive
and are shown in Figure \ref{IRseds}.  Temperatures of 800 and 850 K for the hot
dust components in SSCs 1 and 2 respectively, are effective at reproducing the
observed near-IR SEDs.  \citet{Vanzi00} also concluded that hot dust
was necessary to explain the red $H-K$ colors they observed in SBS 0335-052.

\section{Evidence for Successive Cluster Formation in SBS 0335-052}\label{suc_sec}

We have already established a significant age spread between the SSCs in
SBS 0335-052 (\S\ref{agesec}).  In this section, we investigate the possibility
of successive cluster formation in the galaxy. 
\citet{Thuan97} demonstrated a systematic increase in the [F569W]$-$[F791W]
color away from SSC 1.  They attributed this trend to a combination of
differential extinction by dust and evolutionary effects arising from
sequential (self) propagating star formation.  \citet{Thompson06} also
suggest that each star-formation event has positive feedback on subsequent
star formation.

\begin{figure}
\begin{center}$
\begin{array}{cc}
\includegraphics[width=3.4in]{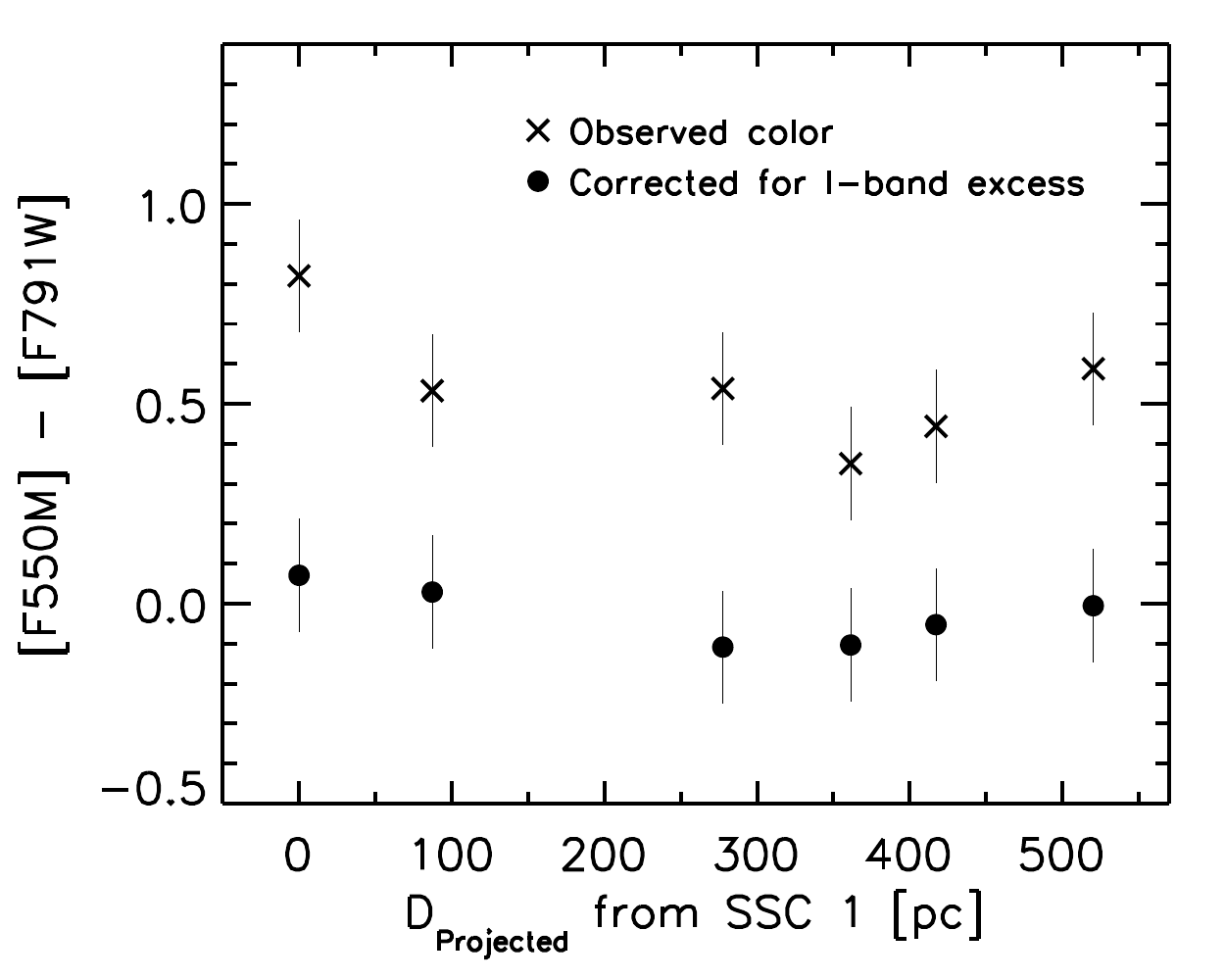} \\
\includegraphics[width=3.4in]{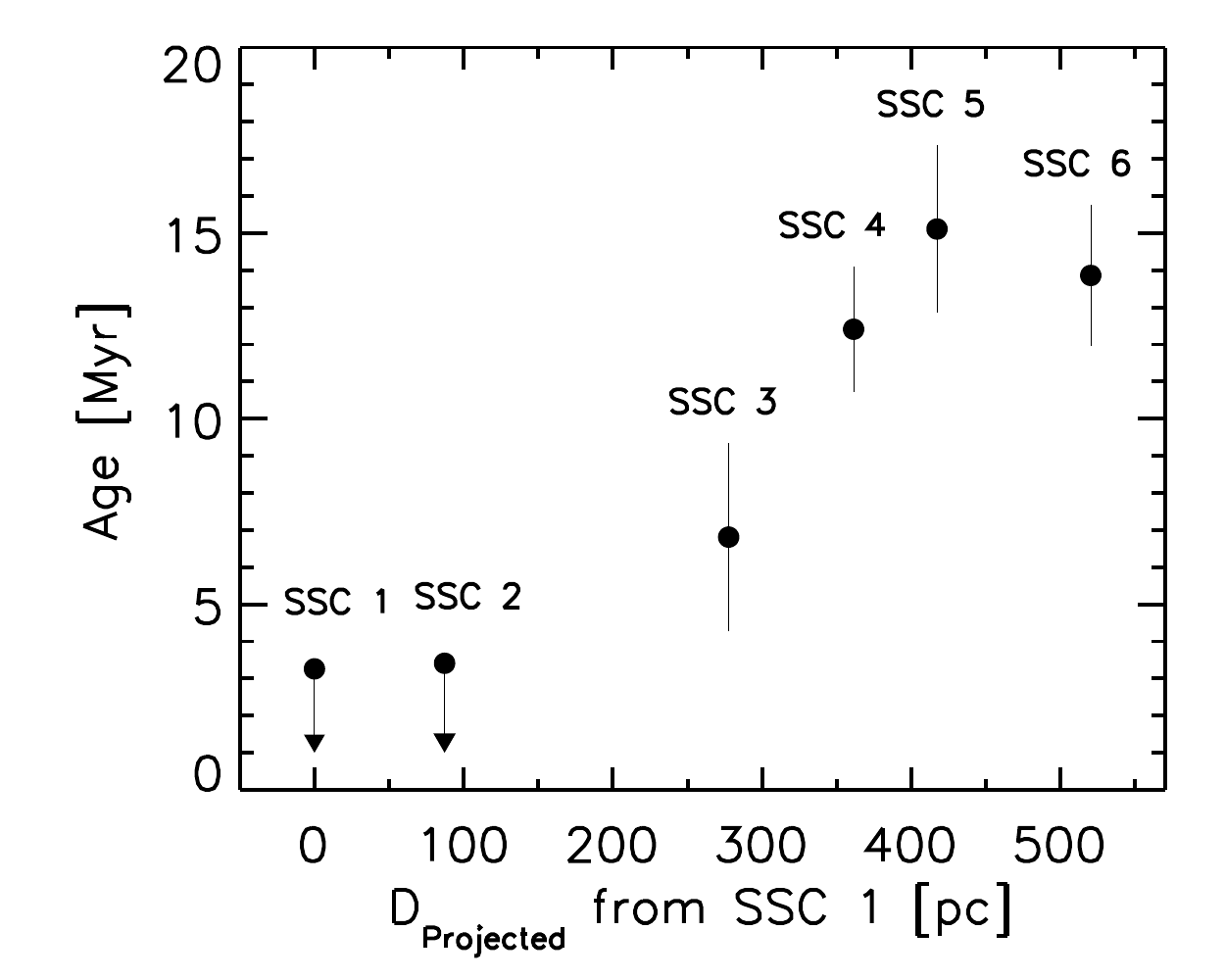}
\end{array}$
\caption{{\it Top:} [F550M]$-$[F791W] ($V-I$) color versus projected distance
from SSC 1.  Observed colors and colors corrected for the $I$-band excess
(Table \ref{excess}) are shown.  We do not observe a
correlation between $V-I$ and distance, as \citet{Thuan97} did. 
This is due to the different $V$-band filters used: the F569W filter used
by \citet{Thuan97} is highly contaminated by gaseous emission lines, whereas
the F550M filter used here is not.
{\it Bottom:} Age (from H$\alpha$ equivalent width) versus projected distance from SSC 1.
There is a strong
correlation indicating successive cluster formation in SBS 0335-052
from the northern (SSC 6) to southern (SSC 1) end of the galaxy.
\label{dist_age}}
\end{center}
\end{figure}

In the top of Figure \ref{dist_age}, we plot [F550M]$-$[F791W] color versus
projected distance from SSC 1, similar to Figure 2 in \citet{Thuan97}.
We use the same $I$-band (F791W) filter here, but
a different (line-free) $V$-band filter (F550M).  The F569W filter used
by \citet{Thuan97} contains significant contamination from gaseous
emission lines, whereas the F550M filter does not.  We do not observe
the same strong trend of redder [F550M]$-$[F791W] colors with distance from SSC 1.

We do, however, observe a strong correlation between cluster age (from H$\alpha$
equivalent width) and distance from SSC 1 in the bottom of
Figure \ref{dist_age}.  This is clear evidence for successive
cluster formation from the northern end of SBS 0335-052 (SSC 6)
to the southern end of the galaxy (SSC 1): we calculate a ``propagation''
speed of $\sim 35$ pc Myr$^{-1} \approx 35$ km s$^{-1}$.  It should be emphasized
that successive cluster formation does not necessarily imply self-propagating
cluster formation (i.e. SSCs triggering the formation of more SSCs through
supernovae shocks), although this is a possibility.  Alternatively,
the SSCs in SBS 0335-052 may have been sequentially triggered by
a large-scale disturbance sweeping through the galaxy from roughly north
to south, the origin of which is presently unknown, but could be related to an
interaction with neighboring galaxies SBS 0335-052W and NGC 1376 \citep{Pustilnik01}.

\section{Conclusions}\label{conclusions}

We have presented a multi-wavelength study of the super star clusters in
SBS 0335-052, the lowest metallicity galaxy with a star formation rate
$\gtrsim 1~M_\odot$~yr$^{-1}$ known.  We use new near-IR and archival optical {\it HST}
observations as well as free-free radio continuum measurements \citep{Johnson08}
of the galaxy to probe the stellar populations and the
gas and dust components of the clusters.  The main results of our work are summarized below:

\begin{enumerate}

\item{There is a significant age spread between the SSCs in SBS 0335-052
that is correlated with position in the galaxy.
The ages are in the range $\lesssim 3$ Myr to $\sim 15$ Myr with the youngest clusters
located in the south and the oldest clusters in the north.  A large-scale
disturbance, with a velocity of $\sim 35$ km s$^{-1}$,
appears to have triggered the successive cluster formation that we observe.}

\item{We find evidence for a porous and clumpy ISM surrounding the youngest SSCs
in SBS 0335-052 (1 and 2).  The measured ionizing luminosities from H$\alpha$,
Pa$\alpha$, {\it and} optically-thin, free-free radio emission \citep{Johnson08}
are lower than expected compared to the optical SEDs, suggesting a fraction of ionizing
photons from the stellar continuum are escaping the strict confines of the clusters before
ionizing hydrogen, that would have otherwise contributed to the measured ionized gas emission
(i.e. H$\alpha$, Pa$\alpha$, and radio).  The $A_V$'s of SSCs 1 and 2 derived from SED fitting
($\sim 0.5$ mag) also indicate the existence of low-extinction regions that provide
relatively unobscured lines-of-sight into the embedded clusters.  The corrected, intrinsic
ionizing luminosities of SSCs 1 and 2 are each $\sim 5 \times 10^{52}~{\rm s}^{-1}$ (the
equivalent of $\sim 5000$ O7.5 V stars) and they each have total stellar masses
of $\sim 10^6 M_\odot$.}

\item{An $I$-band (F791W, $\sim 0.8~\mu$m) excess with respect to model SEDs is observed for all of the
SSCs in SBS 0335-052.  The $I$-band magnitudes are $\sim 0.5-0.7$ mag brighter than the
model predictions.  A similar result was found in the young massive clusters in NGC 4449
\citep{Reines08} and attributed to a photoluminescent process known as Extended Red Emission.
We hypothesize that the same mechanism dominates the $I$-band excess in SBS 0335-052.}

\item{All of the SSCs have red near-IR ($\sim 1.6-2.1~\mu$m) colors and large near-IR
excesses with respect to model SEDs fit to the optical photometry.  The red near-IR colors,
however, are quantitatively different (redder) in the youngest SSCs (1 and 2) compared to
the older SSCs (3-6), clearly
indicating a different origin for the red colors.  We have demonstrated that the near-IR
light from the older clusters is dominated by evolved red supergiants, whereas the near-IR
light from the youngest clusters is dominated by hot dust emission.  The young SSCs
1 and 2 are also the only radio-detected clusters in the galaxy \citep{Johnson08}.}

\item{We have proposed a scenario that can account for the apparently discrepant
extinction estimates found in the literature for the starburst region
encompassing the youngest SSCs (1 and 2) in SBS 0335-052.  Our picture is consistent
with all of the emission (at all wavelengths) being associated with the optically visible
SSCs 1 and 2, and is based on our calculated absorptions of the optical and IR recombination
lines relative to the (thermal) free-free radio emission \citep{Johnson08}.
We find that an ISM containing dense dust clumps harbouring large grains ($\sim 1-2~\mu$m),
with a dust clump covering factor of $\sim 60\%$, can account for
the absorption of the optical/near-IR recombination lines relative to the radio, as well as the
large extinctions derived from previous mid-IR observations.  Interclump regions containing
diffuse dust can account for the low {\it measured} extinctions derived from
optical/near-IR observations.}

\end{enumerate}

The main goal of this case study was to investigate the formation of
massive star clusters in an extremely low metallicity environment
that might be analogous to conditions in which the now ancient
globular clusters were born.  In general, the formation of massive
star clusters in SBS 0335-052 appears to be similar to that found in
higher metallicity counterparts, e.g. Henize~2-10 and Haro~3
\citep{Johnson03, Johnson04}.  These galaxies are all blue compact dwarfs
(BCDs), and each of them currently hosts radio-detected massive star clusters
with ages $\lesssim 3$~Myr and masses of $\sim 10^6 M_\odot$, all of which are
consistent with evolving into globular clusters in several billion
years.  Perhaps even the low metal abundance in SBS 0335-052 is insufficiently extreme
to reflect primordial star formation; in fact, models predict that the transition
from metal-free massive Population III stars to ``normal'' (less massive) Population II
occurs at $Z/Z_\odot\lesssim 10^{-5}$ \citep{Mackey03,Bromm04}.  On the other hand,
large metallicity fluctuations because of radiative feedback effects
may enable ``hidden'' Population III star formation at slightly
higher metal abundances \citep{Tornatore07}.  Clearly more detailed studies
need to be carried out on metal-poor star-forming galaxies in the Local Universe,
and comparative samples need to be obtained in order to investigate
interactions and trends among star cluster and dust formation,
environment, and metal enrichment.

\acknowledgments

We thank the anonymous referee for numerous comments and suggestions that
improved the overall quality of the paper.  A.E.R. appreciates useful
discussions with David Nidever, Ricardo Schiavon and Remy Indebetouw, and
is grateful for support from the Virginia Space Grant Consortium.
Support for Program number GO-10894.01-A was provided by NASA
through a grant from the Space Telescope Science Institute, which is
operated by the Association of Universities for Research in Astronomy,
Incorporated, under NASA contract NAS5-26555.  K.E.J. gratefully
acknowledges support for this paper provided by NSF through CAREER award
0548103 and the David and Lucile Packard Foundation through a Packard
Fellowship.



\begin{thebibliography}{}

\bibitem[Brodie \& Strader(2006)]{Brodie06} Brodie, J.~P., \& Strader, J.\ 2006, \araa, 44, 193 

\bibitem[Bromm \& Larson(2004)]{Bromm04} Bromm, V., \& Larson, R.~B.\ 2004, \araa, 42, 79

\bibitem[Bromm \& Loeb(2003)]{Bromm03} Bromm, V., \& Loeb, A.\ 2003, \nat, 425, 812 

\bibitem[Calzetti et al.(2000)]{Calzetti00} Calzetti, D., Armus, 
L., Bohlin, R.~C., Kinney, A.~L., Koornneef, J., \& Storchi-Bergmann, T.\ 
2000, \apj, 533, 682

\bibitem[Cardelli et al.(1989)]{Cardelli89} Cardelli, J.~A., 
Clayton, G.~C., \& Mathis, J.~S.\ 1989, \apj, 345, 245 

\bibitem[Condon(1992)]{Condon92} Condon, J.~J.\ 1992, \araa, 30, 575 

\bibitem[Dale et al.(2001)]{Dale01} Dale, D.~A., Helou, G., 
Neugebauer, G., Soifer, B.~T., Frayer, D.~T., 
\& Condon, J.~J.\ 2001, \aj, 122, 1736 

\bibitem[Darbon et al.(1998)]{Darbon98} Darbon, S., Perrin, J.-M., \&
Sivan, J.-P.\ 1998, \aap, 333, 264 

\bibitem[Fall \& Rees(1985)]{Fall85} Fall, S.~M., \& Rees, M.~J.\ 1985, \apj, 298, 18 

\bibitem[Fall \& Zhang(2001)]{Fall01} Fall, S.~M., \& Zhang, Q.\ 2001, \apj, 561, 751 

\bibitem[Fitzpatrick(1985)]{Fitzpatrick85} Fitzpatrick, E. L. 1985, \apj, 299, 219

\bibitem[Fitzpatrick \& Massa(1990)]{Fitzpatrick90} Fitzpatrick, 
E.~L., \& Massa, D.\ 1990, \apjs, 72, 163 

\bibitem[Freeman \& Bland-Hawthorn(2002)]{Freeman02} Freeman, K., \& Bland-Hawthorn,
J.\ 2002, \araa, 40, 487 

\bibitem[Fruchter \& Hook(2002)]{Fruchter02} Fruchter, A.~S., \& Hook, R.~N.\ 2002,
\pasp, 114, 144 

\bibitem[Gordon et al.(1997)]{Gordon97} Gordon, K.~D., Calzetti, D., \&
Witt, A.~N.\ 1997, \apj, 487, 625

\bibitem[Harris(1991)]{Harris91} Harris, W.~E.\ 1991, \araa, 29, 543 

\bibitem[Houck et al.(2004)]{Houck04} Houck, J.~R., et al.\ 
2004, \apjs, 154, 211 

\bibitem[Hunt et al.(2005)]{Hunt05} Hunt, L., Bianchi, S., \&
Maiolino, R.\ 2005, \aap, 434, 849

\bibitem[Hunt et al.(2004)]{Hunt04} Hunt, L.~K., Dyer, K.~K., Thuan, T.~X., \&
Ulvestad, J.~S.\ 2004, \apj, 606, 853

\bibitem[Hunt et al.(2001)]{Hunt01} Hunt, L.~K., Vanzi, L., \& Thuan, T.~X.\ 2001,
\aap, 377, 66 

\bibitem[Ivezic \& Elitzur(1997)]{Ivezic97} Ivezic, Z., \& Elitzur, M.\ 1997,
\mnras, 287, 799 

\bibitem[Ivezic et al.(1999)]{Ivezic99} Ivezic, Z., Nenkova, M., 
\& Elitzur, M.\ 1999, User Manual for DUSTY (Lexington: Dept. Phys. Astron.,
Univ. Kentucky)

\bibitem[Izotov et al.(2001)]{Izotov01} Izotov, Y.~I., Chaffee, F.~H.,
\& Schaerer, D.\ 2001, \aap, 378, L45 

\bibitem[Izotov et al.(1990)]{Izotov90} Izotov, I.~I., Guseva, 
N.~G., Lipovetskii, V.~A., Kniazev, A.~I., 
\& Stepanian, J.~A.\ 1990, \nat, 343, 238 

\bibitem[Izotov et al.(1997)]{Izotov97} Izotov, Y.~I., 
Lipovetsky, V.~A., Chaffee, F.~H., Foltz, C.~B., Guseva, N.~G., 
\& Kniazev, A.~Y.\ 1997, \apj, 476, 698 

\bibitem[Johnson, Hunt, \& Reines(in prep)]{Johnson08} Johnson, K.~E., Hunt,
L.~K., \& Reines, A.~E. \ 2008, in prep

\bibitem[Johnson et al.(2004)]{Johnson04} Johnson, K.~E., 
Indebetouw, R., Watson, C., \& Kobulnicky, H.~A.\ 2004, \aj, 128, 610 

\bibitem[Johnson \& Kobulnicky(2003)]{Johnson03} Johnson, K.~E., \&
Kobulnicky, H.~A.\ 2003, \apj, 597, 923 

\bibitem[Kennicutt(1998)]{Kennicutt98} Kennicutt, R.~C., Jr.\ 1998, \araa, 36, 189 

\bibitem[Kim et al.(1994)]{Kim94} Kim, S.-H., Martin, P.~G., 
\& Hendry, P.~D.\ 1994, \apj, 422, 164 

\bibitem[Krist \& Hook(2004)]{Krist04} Krist, J., \& Hook, R.\ 2004,
The Tiny Tim User’s Guide (Baltimore: STScI)

\bibitem[Kroupa(2001)]{Kroupa01} Kroupa, P.\ 2001, \mnras, 322, 231

\bibitem[Leitherer et al.(1999)]{Leitherer99} Leitherer, C., et 
al.\ 1999, \apjs, 123, 3 

\bibitem[Mackey et al.(2003)]{Mackey03} Mackey, J., Bromm, V., \& Hernquist, L.\ 2003,
\apj, 586, 1

\bibitem[Maiolino \& Natta(2002)]{Maiolino02} Maiolino, R., \& Natta, A.\ 2002,
\apss, 281, 233

\bibitem[McCrady et al.(2003)]{McCrady03} McCrady, N., Gilbert, 
A.~M., \& Graham, J.~R.\ 2003, \apj, 596, 240 

\bibitem[Melnick et al.(1992)]{Melnick92} Melnick, J., Heydari-Malayeri, M., \&
Leisy, P.\ 1992, \aap, 253, 16 

\bibitem[Misselt et al.(1999)]{Misselt99} Misselt, K.~A., 
Clayton, G.~C., \& Gordon, K.~D.\ 1999, \apj, 515, 128 

\bibitem[Osterbrock(1989)]{Osterbrock89} Osterbrock, D.~E.,
Astrophysics of Gaseous Nebulae and Active Galactic Nuclei
(Mill Valley: University Science Books)

\bibitem[Peebles \& Dicke(1968)]{Peebles68} Peebles, P.~J.~E., \& 
Dicke, R.~H.\ 1968, \apj, 154, 891 

\bibitem[Perrin \& Sivan(1992)]{Perrin92} Perrin, J.-M., \&
Sivan, J.-P.\ 1992, \aap, 255, 271 

\bibitem[Pickles(1998)]{Pickles98} Pickles, A.~J.\ 1998, \pasp, 
110, 863 

\bibitem[Plante \& Sauvage(2002)]{Plante02} Plante, S., \& Sauvage, M.\ 2002, \aj, 124, 1995 

\bibitem[Pustilnik et al.(2001)]{Pustilnik01} Pustilnik, S.~A., 
Brinks, E., Thuan, T.~X., Lipovetsky, V.~A., 
\& Izotov, Y.~I.\ 2001, \aj, 121, 1413 

\bibitem[Reines et al.(2008)]{Reines08} Reines, A.~E., Johnson, 
K.~E., \& Goss, W.~M.\ 2008, \aj, 135, 2222 

\bibitem[Santoro \& Shull(2006)]{Santoro06} Santoro, F., \& Shull, J.~M.\ 2006,
\apj, 643, 26 

\bibitem[Schlegel et al.(1998)]{Schlegel98} Schlegel, D.~J., 
Finkbeiner, D.~P., \& Davis, M.\ 1998, \apj, 500, 525 

\bibitem[Thompson et al.(2006)]{Thompson06} Thompson, R.~I., 
Sauvage, M., Kennicutt, R.~C., Jr., Engelbracht, C.~W., 
\& Vanzi, L.\ 2006, \apj, 638, 176 

\bibitem[Thuan et al.(1997)]{Thuan97} Thuan, T.~X., Izotov, 
Y.~I., \& Lipovetsky, V.~A.\ 1997, \apj, 477, 661 

\bibitem[Thuan et al.(1999)]{Thuan99} Thuan, T.~X., Sauvage, 
M., \& Madden, S.\ 1999, \apj, 516, 783 

\bibitem[Tornatore et al.(2007)]{Tornatore07} Tornatore, L., Ferrara, A., \&
Schneider, R.\ 2007, \mnras, 382, 945

\bibitem[Tumlinson(2007)]{Tumlinson07} Tumlinson, J.\ 2007, \apj, 
665, 1361 

\bibitem[Vacca et al.(1996)]{Vacca96} Vacca, W.~D., Garmany, 
C.~D., \& Shull, J.~M.\ 1996, \apj, 460, 914 

\bibitem[Vandenberg et al.(1996)]{Vandenberg96} Vandenberg, D.~A., Stetson, P.~B.,
\& Bolte, M.\ 1996, \araa, 34, 461 

\bibitem[van Hoof et al.(2004)]{vanHoof04} van Hoof, P.~A.~M., 
Weingartner, J.~C., Martin, P.~G., Volk, K., 
\& Ferland, G.~J.\ 2004, \mnras, 350, 1330 

\bibitem[Vanzi et al.(2000)]{Vanzi00} Vanzi, L., Hunt, L.~K.,
Thuan, T.~X., \& Izotov, Y.~I.\ 2000, \aap, 363, 493 

\bibitem[Whitmore(2003)]{Whitmore03} Whitmore, B.~C.\ 2003, A 
Decade of Hubble Space Telescope Science, 153 

\bibitem[Whitmore et al.(2007)]{Whitmore07} Whitmore, B.~C., 
Chandar, R., \& Fall, S.~M.\ 2007, \aj, 133, 1067 

\bibitem[Whitney et al.(2003)]{Whitney03} Whitney, B.~A., Wood, 
K., Bjorkman, J.~E., \& Wolff, M.~J.\ 2003, \apj, 591, 1049 

\bibitem[Witt \& Vijh(2004)]{Witt04} Witt, A.~N., \& Vijh, U.~P.\ 2004,
Astrophysics of Dust, 309, 115 

\end{thebibliography}
\end{document}